\documentclass[onefignum,onetabnum]{siamonline220329}

\usepackage{lipsum}
\usepackage{amsfonts}
\usepackage{graphicx}
\usepackage{epstopdf}
\usepackage{algorithmic}
\usepackage{amsmath,amssymb,bm}
\ifpdf
  \DeclareGraphicsExtensions{.eps,.pdf,.png,.jpg}
\else
  \DeclareGraphicsExtensions{.eps}
\fi

\usepackage{enumitem}
\setlist[enumerate]{leftmargin=.5in}
\setlist[itemize]{leftmargin=.5in}

\newsiamremark{remark}{Remark}
\newsiamremark{hypothesis}{Hypothesis}
\crefname{hypothesis}{Hypothesis}{Hypotheses}
\newsiamthm{claim}{Claim}

\headers{MSCT Reconstruction with Diffusion Priors and INR}{
Shen, Zhang, Dong, Qiu, Cui, and Li
}

\title{Incomplete Data Multi-Source Static Computed Tomography Reconstruction with Diffusion Priors and Implicit Neural Representation\thanks{Submitted to the editors DATE.
}}

\author{Ziju Shen\thanks{School of Mathematical Sciences, Peking University, Beijing 100871, P.R. China (\email{zjshen@pku.edu.cn}).}
\and 
Haimiao Zhang\thanks{Corresponding author. Institute of Applied Mathematics, College of Computer Science, Beijing Information Science and Technology University, Beijing 102206, P.R. China 
  (\email{hmzhang@bistu.edu.cn}
). This author was supported by the National Natural Science Foundation of China (Grant No. 12101061) and the Young Elite Scientist Sponsorship Program By
BAST (No. BYESS2024263).}
\and Bin Dong\thanks{Beijing International Center for Mathematical Research, Peking University, Beijing 100871, P.R. China (\email{dongbin@math.pku.edu.cn}). This author was supported by the New Cornerstone Investigator Program.}
\and Jun Qiu\thanks{Institute of Applied Mathematics, College of Computer Science, Beijing Information Science and Technology University, Beijing 102206, P.R. China (\email{qiu.jun.cn@ieee.org}). This author was supported by the National Natural Science Foundation of China (Grant No. ~61931003).}
\and Yunxiang Li\thanks{Nanovision Technology (Beijing) Co., Ltd., Beijing 100094, P.R. China 
  (\email{yunxiang.li@nanovision.com.cn}, \email{zhili.cui@nanovision.com.cn}
).}
\and Zhili Cui\footnotemark[6]
}

\usepackage{amsopn}

\ifpdf
\hypersetup{
  pdftitle={Haimiao's Article},
  pdfauthor={Z. Shen, H. Zhang, B. Dong, J. Qiu, Y. Li, and Z. Cui}
}
\fi

\graphicspath{{figures/}}

\newcommand{\subfigimg}[3][,]{%
  \setbox1=\hbox{\includegraphics[#1]{#3}}
  \leavevmode\rlap{\usebox1}
  \rlap{\hspace*{1pt}\raisebox{\dimexpr\ht1-0.6\baselineskip}{#2}}
  \phantom{\usebox1}
}

\usepackage{multirow}
\usepackage{makecell}

\begin{document}
\maketitle

\begin{abstract}
The dose of X-ray radiation and the scanning time are crucial factors in computed tomography (CT) for clinical applications. In this work, we introduce a multi-source static CT imaging system designed to rapidly acquire sparse view and limited angle data in CT imaging, addressing these critical factors. This linear imaging inverse problem is solved by a conditional generation process within the denoising diffusion image reconstruction framework. The noisy volume data sample generated by the reverse time diffusion process is projected onto the affine set to ensure its consistency to the measured data. To enhance the quality of the reconstruction, the 3D phantom's orthogonal space projector is parameterized implicitly by a neural network. Then, a self-supervised learning algorithm is adopted to optimize the implicit neural representation. Through this multistage conditional generation process, we obtain a new approximate posterior sampling strategy for MSCT volume reconstruction. Numerical experiments are implemented with various imaging settings to verify the effectiveness of our methods for incomplete data MSCT volume reconstruction.
\end{abstract}



\section{Introduction}
Computed tomography (CT) is widely used in non-destructive industry detection, archaeology, clinics, etc. The three-dimensional X-ray CT (3DCT) is a technique to scan the object and then reconstruct the object with a numerical algorithm to represent the spatial domain 3D inner structure. For clinical applications of CT, there are two kinds of challenges in the data measurement and volume data reconstruction. The first one is the so-called As Low As Reasonably Achievable (ALARA) principle \cite{strauss2006alara}. It means that the object scanning process needs to be controlled to reduce the radiation dose while the volume data should be recovered with as high quality as possible. However, the decrease in measured data will lead to a more seriously ill-posed inverse problem. Therefore, radiation dose reduction requirements lead to more challenging problems in mathematical modeling and numerical algorithm design for 3DCT imaging. Low dose, limited angle, and sparse view X-ray scanning are representative protocols to reduce the X-ray radiation dose during projection measurement. Another challenge of 3DCT in clinic applications is how to speed up the scanning process. The patients' movement or physiological processes, such as cardiac beating during X-ray scanning, will lead to motion artifacts in the reconstructed volume data (or phantom). Hardware equipment upgrade is one of the solutions to accelerate the measurement process. In this work, we study a new 3DCT imaging system with multiple X-ray sources static CT (MSCT) equally distributed around the rotation circle trajectory. This new MSCT system innovation will significantly accelerate X-ray scanning with a new low-radiation dose scanning protocol.

The 3DCT volume reconstruction methods can be classified into two groups: the classical physics-driven image reconstruction algorithms and the deep learning models. The classical methods follow the imaging geometry and physics-driven modeling paradigm. Representative methods are the analytical reconstruction algorithm (i.e., FDK \cite{FDK-1984}) and the regularization model based iterative reconstruction algorithms \cite{sidky2008image}. 
Hyperparameter choosing and numerical algorithm convergence analysis are common topics in the classical incomplete data 3DCT volume reconstruction methods. 
The deep learning based approach is mainly focused on data-driven modeling. The large-scale training set is required for deep neural network (DNN) weight optimization. The neural network architecture and learning framework design are popular topics for high-quality incomplete data 3DCT volume reconstruction \cite{zhang2020review,wang2020DLCTnature}. The main issues in the DNN-based volume reconstruction approaches are model generalization and explainability. The following subsections present a more detailed discussion of the two kinds of 3DCT volume reconstruction methods.

\subsection{Physics-Driven 3DCT Reconstruction Approaches}\label{sec:physics-driven-models-review}
The classical 3DCT volume reconstruction algorithms are the analytical reconstruction algorithms, algebraic reconstruction technique (ART), and optimization models. These volume reconstruction approaches are designed based on the imaging physics and thus called physics-driven modeling. The representative analytical volume reconstruction algorithm is the so-called FDK algorithm proposed by L. A. Feldkamp, L. C. Davis, and J. W. Kress \cite{FDK-1984}, which is the 3D filtered backprojection algorithm (FBP). The FDK algorithm is convenient to implement and produces high-quality volume reconstruction results when the number of scanning views is large enough to satisfy the sampling theorem requirements. The ART algorithm \cite{gordon1974ART} and its variant, simultaneous ART (SART) \cite{jiang2003convergenceSART}, is a kind of iterative reconstruction algorithm that can be deduced from the optimization model. Optimization models for sparse view 3DCT imaging tasks are more promising in producing high-quality reconstruction than the FDK and ART algorithms. From the mathematical modeling viewpoint, the optimization model provides a convenient framework to incorporate the imaging physics and the data priors in the reconstruction process. Representative image priors are the total variation \cite{rudin1992ROF}, wavelet frames \cite{dong2010mra}, dictionary learning \cite{bai2017CTDictLearn,xu2012XrayDictL}, and low rank tensor norm \cite{cai2014CineCBCT}. Imaging physics provides knowledge on how to constrain the reconstructed volumetric data to be consistent with the measured data. This is also called the data fidelity term in the optimization model. Based on the compressed sensing theory \cite{donoho2006compressed,eldar2012compressed}, we can obtain a high-precision solution to the linear imaging inverse problem by properly designed objective functional and numerical algorithms. The main drawback of the iterative reconstruction algorithm is the trial-and-error strategy for hyperparameters fine-tuning for each reconstructed phantom. The convergence behavior to the optimal solution is another challenge of the iterative reconstruction algorithm. The computation time of the iterative algorithms is often longer than the analytic reconstruction approaches.

\subsection{Data-Driven 3DCT Reconstruction Approaches}
Deep learning is a widely used data-driven modeling approach that has applications in natural language processing, computer vision, and audio processing \cite{lecun2015deep,bengio2017deep}. For CT imaging, the modern deep learning models can be summarized into two classes:(1) the non-generative models and (2) the generative models. For the non-generative model, the DNN is trained to approximate the mapping between the measured data (or projection) to the imaging object (i.e., 2D image or 3D volumetric data) \cite{zhu2018AutoMap}. The deep learning models are learned in an end-to-end manner. For the generative models, a common way to design the image reconstruction model is to adopt a pre-trained DNN as the image priors and plug it into a physics-driven model. This is the so-called plug-and-play image processing framework \cite{venkatakrishnan2013PnP,zhang2021DNNdenoiser}. In the following, we will dive deep into these deep learning models for 3DCT volume reconstruction.

\subsubsection{Non-generative Models for CT Reconstruction}
The recently proposed unrolled dynamics (UD) models are a new mathematical modeling methodology that combines traditional image reconstruction (or restoration) models with DNN \cite{zhang2020review,monga2021AlgUnroll}. For example, some of the components of the iterative image reconstruction algorithms are approximated or replaced by neural network modules. The proximal operators in iterative CT image reconstruction algorithms, i.e., primal-dual algorithm \cite{chambolle2011PrimalDualAlg,zhang2011PrimalDualAlg}, are approximated by neural network modules \cite{adler2018PDNet}. For an iterative image restoration task, the image denoising step is replaced by a learned DNN denoiser \cite{zhang2021DNNdenoiser}. The hyperparameters in the soft-thresholding operation of the alternating direction method of multipliers (ADMM) algorithm \cite{boyd2011ADMM} are learned by a neural network module \cite{yang2016admmNet,yang2018admmCSnet}. For 2D sparse view CT, the authors in \cite{zhang2020metainv} proposed to predict the variables initialization of conjugate gradient (CG) algorithm by a DNN. The adopted DNN module works as a hypernetwork \cite{ha2017hypernetworks} to build the unrolled half quadratic splitting (HQS) algorithm \cite{geman1995HQS} based neural network architecture. For general incomplete data CT image reconstruction tasks, authors in \cite{zhang2019jsr,wang2021LACTmiccai,wang2021indudonet,wang2023indudonet+,wang2022dudotrans} showed that the UD-based deep learning models have better generalization and explainability than the pure DNN models.

\subsubsection{Generative Models for CT Reconstruction}
Generative models such as the generative adversarial networks (GANs) \cite{goodfellow2014generative}, variational autoencoder (VAE) \cite{kingma2013VAE}, and diffusion models are representative methods of data-driven image prior. These models are created for unconditional new text, image, and video generation. Recently, generative models have been utilized as a powerful image prior and are introduced in low-level vision tasks such as image restoration (i.e., inpainting, denoising, and super-resolution) and image reconstruction (i.e., CT and MRI). 

For conditional image synthesis, the generative models need guidance information to control the content and semantics of the generated images. The class labels can guide the GANs model to generate specific classes of images \cite{dhariwal2021diffusion}. In diffusion process-based image generation models, the score function and likelihood function are used to guide the conditional generation process. In practice, the GANs model has a faster inference than the diffusion model. However, recent works show that the diffusion model has more excellent image generation performance at various tasks than GANs \cite{dhariwal2021diffusion,rombach2022high}.

In diffusion-based conditional generation models, the main challenge is how to tackle the posterior sampling problem. That is, how can we sample from $p_t(\bm{x}|\bm{y})$ to obtain a sample $\bm{x}_t$ while it is restricted to be consistent with the measured data $\bm{y}$ at the time step $t$? 
Previous works on conditional diffusion generation for imaging tasks can be categorized into three classes: (1) adding a data consistency constraint in each reverse time diffusion step of the unconditional diffusion sampling process \cite{chung2022MCG,chung2023DPS,song2021scoreSDE,chung2022accMRI}, (2) guiding the reverse time diffusion steps by an estimated conditional score function  \cite{chung2023DPS,dou2024diffusion,song2023pseudoinverse}, (3) training an additional neural network to guide the sampling of Bayes posterior. The first class of approaches is somewhat ad hoc because it uses a projection operation to restrict the generated sample to be consistent with the measured data. The second class of approaches is derived based on the Bayesian law such that the gradient of the posterior $p(x|y)$ is estimated by the gradients of both the likelihood $p(y|x)$ and the image prior $p(x)$ at each time step. The aforementioned methods are challenging to implement for inverse problems with noisy measurement and nonlinear imaging processes or fail to produce desirable reconstructed images. The third class of approaches needs to train a new network for each task. Therefore, it is impractical for the CT reconstruction tasks at various scanning protocols. 

This work introduces a new conditional diffusion model for the MSCT system. The diffusion image prior (DIP) is utilized in the volume reconstruction process to model each slice of the phantoms. 
An affine set projection operation is introduced to restrict the reverse time generated sample to be consistent with the measured data. In practice, a high spatial resolution 3D phantom is desired, so we proposed to utilize the implicit neural representation (INR) model to represent the reconstructed volumetric data. To suppress the accumulated error in the reverse time generation process, the measured data is used as supervision within a self-supervised learning (SSL) model to refine the reconstructed phantom further. In summary, our newly proposed MSCT volume reconstruction approach combines the diffusion based image prior, affine set projection constraint, INR, and SSL techniques. We denote the newly proposed model as DIP-ASPINS. The proposed DIP-ASPINS model was tested on incomplete data (i.e., sparse view and limited angle) MSCT imaging tasks under different system settings to verify its effectiveness.

The paper is organized as follows. \Cref{sec:related-work} reviews the related works and backgrounds of our methods. The newly proposed MSCT volume reconstruction approaches and algorithm summarization are presented in \Cref{sec:our-methods}. 
The experimental results on the simulated MSCT data are reported in \Cref{sec:experiments}. The conclusions and future work are outlined in \Cref{sec:conclusions}.

\section{Related Works}\label{sec:related-work}
\subsection{Classical Reconstruction Model}
The 3DCT imaging problem is a linear inverse problem that the following form can define
\begin{equation}\label{eq:linear-system}
    \bm{Y}=\bm{Pu}+ \bm{n},
\end{equation}
where $\bm{Y}$ is the CT imaging system's measured data (or projection). $\bm{P}$ denotes the forward projection operator, which models the system's physical imaging process. $\bm{u}$ is the volume data to be reconstructed, $\bm{n}\sim \mathcal{N}(0,\sigma \bm{I})$ denotes the additive white Gaussian noise (AWGN) and $\sigma>0$ is the noise level. Note that the noise is essentially a mixture of Gaussian (electronic noise) and Poisson noise in low-dose CT imaging. Here, we adopt a linear system to simplify the presentation.

For the short scan setting of the 3DCT imaging task, the unknown variables in $\bm{u}$ are usually more than the measured data voxels. Thus, the linear system is often undetermined, which leads to an ill-posed inverse problem. To restrict the solution subspace, a regularized optimization model (or variational model) is commonly adopted with the following form
\begin{equation}\label{eq:L2+Ru}
    \min_{\bm{u}}\|\bm{Pu}-\bm{Y}\|^2+\lambda R(\bm{u}),
\end{equation}
where $\lambda \in \mathbb{R}^{++}$ (real positive value) is the regularization parameter, and $R(\bm{u})$ is the regularization term to reflect the prior distribution assumption of $\bm{u}$. The example assumptions on the solution $\bm{u}$ penalize the sparse representation of $\bm{u}$ in the transform domain or the smoothness of $\bm{u}$ in the spatial domain. When the regularization term in \eqref{eq:L2+Ru} is chosen as the indicator function $R(x)=0 $ if $x \in \mathbb{R}^{+}$ (non-negative value) and $R(x)=+\infty $ otherwise, it is denoted as an L2 model. When $R(\bm{u})$ is chosen as the total variation regularization \cite{rudin1992ROF}, the model \eqref{eq:L2+Ru} is denoted as an L2TV model.
In the numerical algorithm, the proximal operator of $R(\cdot)$ is often chosen as a thresholding operator to prompt the sparsity of the transform domain coefficients of the reconstructed 3D volume data. In the plug-and-play (PnP) modeling philosophy \cite{venkatakrishnan2013PnP}, the proximal operator of $R(\cdot)$ can also be replaced directly by an available off-the-shelf image denoiser, such as BM3D \cite{dabov2007BM3D}, NLM \cite{Buades2005NLM}, deep image prior \cite{ulyanov2018DIP}, or Denoising CNN (DnCNN) \cite{zhang2017DnCNN}, to ensure the regularization effect \cite{RED-2017-SIAM}. In this work, we will build our 3DCT volume reconstruction model referring to the foundation model \eqref{eq:L2+Ru}.

\subsection{Implicit Neural Representation (INR) of the 3D Phantom}\label{sec:INR-background}
Volume data in 3DCT imaging is usually represented as a 3D voxel grid in the spatial domain. If volume data are continuously represented in the spatial domain, it is convenient to re-sample the reconstructed image slices to different resolutions.
In order to represent the volume data $\bm{u}$ in a continuous domain, it can be reparameterized by
$$\bm{u}=\mathcal{F}(\bm{\epsilon}_0; \bm{\Phi}),$$
where $\mathcal{F}(\cdot;\cdot)$ is a neural network with parameters $\bm{\Phi}$ and the phantom is encoded by a code tensor  $\bm{\epsilon}_0$. In practice, $\bm{\epsilon}_0$ can be chosen as the random Gaussian white noise as $\bm{\epsilon}\sim \mathcal{N}(0,I)$, the 3D spatial position (or 3D mesh grid) \cite{mildenhall2021nerf}, or the hash code \cite{muller2022instantNGP}. These implicit neural representation methods adopted position embedding (PE) to represent the continuous volume data. PE is widely used as a new and powerful image representation strategy in computer graphics and vision tasks \cite{mildenhall2021nerf,muller2022instantNGP} for novel view synthesis and 3D scene representation. 

In this work, we adopt the PE to represent the 3D volume data in MSCT. For the 3DCT imaging problem \eqref{eq:linear-system}, a self-supervised learning model can be constructed to reconstruct the phantom $\bm{u}$ from the measured projection $\bm{Y}$ by the following optimization problem 
\begin{equation}\label{eq:INR-optimization-Model}
    \min_{\bm{\Phi}}\|\bm{P}\mathcal{F}(\bm{\epsilon}_0; \bm{\Phi})-\bm{Y}\|^2+\lambda \tilde{R}(\bm{\Phi}),
\end{equation}
where the first term provides a data consistency constraint. $\bm{P}$ is the forward projection operator in 3DCT that is usually computed by ray-driven or pixel-driven approaches \cite{van2016ASTRA}. $\bm{Y}$ is the measured projection data by the X-ray scanning equipment. $\bm{\epsilon}_0$ is chosen as the 3D mesh grid in the spatial domain of a unit cube $[0,1]^3$. An alternative way to represent the forward projection operation is to compute the projection in each detector bin by volume rendering to separately model the X-ray attenuation process \cite{zha2022naf}.
The second term of the objective functional in \eqref{eq:INR-optimization-Model} with a balancing hyperparameter $\lambda>0$ is added to regularize the training stability and the parameter distribution \cite{loshchilov2019decoupled,hanson1988comparing}. Note that if $\tilde{R}(\bm{\Phi})$ is chosen as the $L_2$ weight regularization, it is equivalent to the weight decay setting of an optimizer only in special cases \cite{loshchilov2019decoupled}. The authors in \cite{aitchison2018unified} provide a Bayes filtering perspective on the stochastic optimization algorithm with weight decay. 

When the model in \eqref{eq:INR-optimization-Model} is trained with the optimal parameters $\bm{\Phi}^\ast$, the continuously represented 3D volume data can be obtained by $\hat{\bm{u}}=\mathcal{F}(\bm{\epsilon}_0, \bm{\Phi}^\ast)$.
We will adopt the INR-based continuous volume representation in Section \ref{sec:our-methods} for the proposed 3DCT image reconstruction approaches.

\subsection{Diffusion Model} Diffusion models provide two opposite processes to describe the transitions between data distribution and noise distribution. The forward process models how a data point from a prescribed dataset underlying some distribution is transitioned to random noise. The reverse or generative process describes how a data sample is gradually refined from noise or an implicit embedding. Due to the powerful modeling ability of the diffusion model for various modality of dataset, it has been used in various inverse problems in medical imaging \cite{du2024dper,jalal2021robust,chung2022accMRI,song2021solvingIP}, phase retrieval \cite{li2024diffFPR}, natural image restoration \cite{kawar2022DDRM,choi2021ILVR}, and astronomy \cite{karchev2022strong}. In this work, we adopt the diffusion model as an image prior for MSCT image reconstruction. In this subsection, we will review the necessary background on how to define a diffusion model in continuous-time variable and discrete-time variable forms. Then, we will show how existing work adds constraints in the unconditional reverse diffusion model for imaging tasks.

\subsubsection{Continuous Diffusion Models}\label{sec:continuous-diffusion-models}
On the continuous time interval $[0,T]$ with $T>0$, a forward diffusion process  follows the It\^{o} stochastic differential equation (SDE) in the following 
\begin{equation}\label{eq:forward-diffusion-SDE}
{\rm{d}}\bm{x}(t)=\bm{f}(\bm{x}(t),t){\rm{d}}t+g(t){\rm{d}}\bm{w}(t),
\end{equation}
where $\bm{x}(\cdot), \bm{f}(\cdot,\cdot)$ and $\bm{w} \in\mathbb{R}^{n}$. ${\rm{d}}\bm{w}(\cdot)$ represents a ``white noise" and is essentially the derivative of the Wiener process (or Brownian motion) that is independent of $\bm{x}_t=\bm{x}(t)$ \cite{evans2012SDE}. $\bm{f}(\cdot,t)$ and $g(\cdot)$ are the drift and diffusion coefficients, respectively. 

Assume that the latent data distribution of a studied dataset is defined at timestamp $t=0$ as $p_0(\bm{x}_0)$. The continuous distribution $p_t(\bm{x}_t)$ evolves over time according to the SDE \eqref{eq:forward-diffusion-SDE}. It transforms the distribution $p_0(\bm{x}_0)$ into a known simple and tractable distribution $p_T(\bm{x}_T)$ such as the white Gaussian noise $\bm{\epsilon}\sim\mathcal{N}(0,\sigma^2 \bm{I})$ with mean 0 and variance $\sigma$. In this transition path, the sampled data sequence $\bm{x}_t, t\in[0, T]$ starts from a given data sample (i.e., a 3D phantom in 3DCT) $\bm{x}_0\sim p_0(\bm{x}_0)$ is progressively degraded to an almost pure Gaussian noise sample $\bm{x}_T\sim p_T(\bm{x}_T)$ by slowly injecting Gaussian noise \cite{sohl2015deep}. Here, the resulting $\bm{x}_T$ can be viewed as an image embedding in the latent space.

The reverse-time (or backward)  process of \eqref{eq:forward-diffusion-SDE} is also a diffusion process that can be represented by an It\^o SDE of the form
\begin{equation}\label{eq:reverse-diffusion-continuous-variable}
    {\rm{d}}\bm{x}(t)=[\bm{f}(\bm{x}_t,t)-g(t)^{2}\nabla_{\bm{x}_t}\log p(\bm{x}_t)]{\rm{d}}t+g(t){\rm{d}}\bar{\bm{w}},
\end{equation}
where $\bar{\bm{w}}$ is the reverse time Wiener process \cite{anderson1982reverse}. The drift term now depends on the time-related score function $\nabla_{\bm{x}_t}\log p_t(\bm{x}_t)$ which is, in fact, the gradient of the log probability density $p_t(\bm{x}_t)$ with respect to data $\bm{x}_t$. This score function is often intractable and thus is estimated by a neural network $s_{\theta}(\bm{x}_t)=s(\bm{x}_t;\theta)$ with model parameters $\theta$ at each time $t$. In practice, the drift coefficient functional $\bm{f}(\cdot,\cdot)$ and diffusion coefficient function $g(\cdot)$ in \eqref{eq:forward-diffusion-SDE} have different choices \cite{song2021scoreSDE}. When we choose a real-valued function $\beta(t)$ with  $t\in [0,T]$ and set
\begin{align}
    f(\bm{x}_t,t)=-\frac{\beta(t)}{2}\bm{x}_t, \quad  g(t)=\sqrt{\beta(t)},
\end{align}
the forward diffusion model \eqref{eq:forward-diffusion-SDE} is called variance preserving SDE (VP-SDE). The function $\beta(t)>0$ is a time schedule that is often chosen as a linear function over variable $t$. When the drift and diffusion coefficients are chosen as
\begin{equation}
    f(\bm{x}_t,t)=0, g(t)=\sqrt{\frac{\rm{d}\sigma^{2}(t)}{\rm{d}t}},
\end{equation}
where $\sigma(t)$ is the noise level function.
The diffusion model \eqref{eq:forward-diffusion-SDE} is called variance exploding SDE (VE-SDE).

For the specific form of the reverse-time diffusion \eqref{eq:reverse-diffusion-continuous-variable}, VE-SDE can be written as
\begin{equation}\label{eq:VE-SDE}
    {\rm{d}}\bm{x}(t)=\left[-\frac{{\rm{d}}\sigma^{2}(t)}{{\rm{d}}t}\nabla_{\bm{x}_t}\log p(\bm{x}_t)\right]{\rm{d}}t+\sqrt{\frac{{\rm{d}}\sigma^{2}(t)}{{\rm{d}}t}}{\rm{d}}\bar{\bm{w}}(t).
\end{equation}
The data distribution $p_t(\bm{x}_t)$ evolves over the reverse time flow from $t=T$ to $t=0$ following the principle of SDE in \eqref{eq:VE-SDE}. To obtain the generated data sample from the diffusion model, we replace the exact term $\nabla_{\bm{x}_t}\log p(\bm{x}_t)$ by an estimated score function $s_{\theta}(\bm{x}_t)$ and solve the SDE with an adequately designed numerical SDE solver. Therefore, in the sample image generation process, the noise sample (or latent code) $\bm{x}_T\sim p_T(\bm{x}_T)$ is transformed to a data sample $\bm{x}_0\sim p(\bm{x}_0)$  by gradual denoising the sampled data $\bm{x}_t\sim p(\bm{x}_t)$ ($0<t<T$) to the next time stamp data $\bm{x}_{s}$ ($0< s < t$) that following the data distribution $ p_s(\bm{x}_{s}).$  This is an unconditional image generation process.

\subsubsection{Discrete Formulation of the Diffusion Process}
When the time interval $[0,T]$ of the diffusion process is discretized to $N\in\mathbb{N}_{+}$ bins, the diffusion process constructed in denoising diffusion probabilistic models (DDPM) \cite{ho2020DDPM} is described by a Markov chain
$$q(\bm{x}_{1:N}|\bm{x}_0)=\Pi_{k=1}^{N}q(\bm{x}_k|\bm{x}_{k-1})$$
with the transition sequence $\bm{x}_{1:N}=(\bm{x}_1,\bm{x}_2,\cdots,\bm{x}_N)$ starting from $\bm{x}_0$. Each Markov step is a linear Gaussian model
$$q(\bm{x}_k|\bm{x}_{k-1})=\mathcal{N}(\sqrt{\alpha_k}\bm{x}_{k-1},\beta_{k}^2\bm{I})$$
where $\{\alpha_k\}_{k=1}^{N}$ is the noise schedule,
$\beta_k=1-\alpha_k$ is the standard deviation of the noise level. Here, $\bm{I}$ is the unit matrix that reflects the variable correlation. When the time interval $[0,T]$ discretization step $N$ goes to infinity, the Markov chain $\{\bm{x}_{k}\}_{k=1}^{N}$ becomes a continuous stochastic process $\bm{x}_{t}$ with $t\in[0,T]$. The DDPM is equivalent to the VP-SDE in Subsection \ref{sec:continuous-diffusion-models}. The marginal distribution $q(\bm{x}_k|\bm{x}_0)$ can be computed using mathematical induction. For example, the forward transition process is described by
$$\bm{x}_k=\sqrt{\bar{\alpha}_k}\bm{x}_0+\sqrt{1-\bar{\alpha}_k}\cdot \bm{\epsilon}, \quad \bm{\epsilon}\sim \mathcal{N}(0,\bm{I}),$$
where the variables $\bar{\alpha}_k=\Pi_{\ell=1}^{k}\alpha_{\ell}$, $\bm{\epsilon}$ is a random white Gaussian noise. Thus, $\bm{x}_k$ can be calculated analytically with the predefined noise level sequence $\{\alpha_k\}_{k=1}^{N}$, and can be viewed as a noisy version of $\bm{x}_0$. In this diffusion model, $p(\bm{x}_0)$ is the data distribution, and $p(\bm{x}_N)$ is viewed as the prior distribution of the image.

The reverse-time diffusion process is also a Markov chain, and a Gaussian process describes it as
\begin{equation}\label{eq:DDPM-forward}
p_\theta(\bm{x}_{k-1}|\bm{x}_k)=\mathcal{N}(\bm{x}_{k-1};\frac{1}{\sqrt{\alpha_{k}}}(\bm{x}_{k}+(1-\alpha_k)\nabla_{\bm{x}_k}\log p(\bm{x}_k)),(1-\alpha_k)\bm{I}).
\end{equation}
Due to the functional approximation property of the neural network, the score function can be parameterized by a neural network and denoted by $s_{\bm{\theta}}(\bm{x}_{k},\sigma_k)$ with model parameters $\bm{\theta}$ at the noise level $\sigma_k$. The score function can be estimated by training a score-based model with the score matching methods \cite{hyvarinen2005estimation,song2020sliced}.
To find the optimal score function of DDPM, it is trained on a dataset by minimizing  the re-weighted evidence lower bound (ELBO) \cite{ho2020DDPM}
$$\bm{\theta}^{\ast}=\arg\min_{\bm{\theta}}\sum_{k=1}^{N}(1-\alpha_{i})\mathbb{E}_{\bm{x}_0\sim p(\bm{x}_0)}\mathbb{E}_{p(\bm{x}_{k}|\bm{x}_0)}\left[\|s_{\bm{\theta}}(\bm{x}_{k},\sigma_k)-\nabla_{\bm{x}_{k}}\log p(\bm{x}_{k}|\bm{x}_0)\|^{2}_{2}\right].$$

In the image generation process, the first step is to sample a random noise $\bm{x}_N$ from a fixed Gaussian distribution $p(\bm{x}_N)=\mathcal{N}(0,\bm{I})$. Then, based on \eqref{eq:DDPM-forward} and the well-trained unconditional score function $s_{\bm{\theta}^{\ast}}(\cdot,\cdot)$, we can deduce the   data sample $\bm{x}_{k-1}$ from the current $k$-step's sample $\bm{x}_{k}\sim p(\bm{x}_k)$ as 
\begin{equation}\label{eq:DDPM-xk-to-xkm1}
    \bm{x}_{k-1}= \frac{1}{\sqrt{\alpha_k}}\left[\bm{x}_k+(1-\alpha_k)s_{\bm{\theta}^\ast}(\bm{x}_{k},\sigma_k)\right]+\sqrt{1-\alpha_k}\cdot \bm{\epsilon}, \quad \bm{\epsilon} \sim \mathcal{N}(0,I)
\end{equation}
for all the time stamps $k=N,N-1,\cdots,1.$ At last, we obtain a data sample $\bm{x}_0$ that follows the latent image set distribution $p(\bm{x}_0)$.

Note that in the above, both the continuous and discrete form reverse process diffusion models are only constructed for unconditional data generation. To solve widely encountered linear inverse problems in imaging sciences, measured data $\bm{y}$ should be adequately incorporated into the generation process to solve specific imaging tasks. In the following subsection, we will recall existing works on conditional generation utilizing the pre-trained unconditional diffusion models like the ones shown in the above two subsections.

\subsubsection{Conditional Generation Process}\label{sec:conditional-generation-review}
The straightforward approach to obtain a generated image from the diffusion model with a measurement constraint is to train the diffusion model with a conditional score function $\nabla_{\bm{x}}\log p(\bm{x}|\bm{y})$ and use it in the generation process. However, training a task-specific conditional score function for each target imaging inverse problem is time-consuming and unaffordable. A more attractive way to realize conditional generation is by utilizing the pre-trained unconditional generation model as an image prior and by using the measurement to guide the generation process. Existing work proposed heuristic approaches to incorporate the measurement constraint in the generative process can be categorized into two groups: (1) project the unconditionally generated sample to the measurement subspace at each reverse-time diffusion step to meet the measurement constraint; (2) compute the conditional score function approximately. The first class of methods is adopted in applications such as accelerated magnetic resonance imaging (MRI) \cite{chung2022accMRI}, class-conditional image generation \cite{song2021scoreSDE}, super-resolution, and image inpainting \cite{chung2022come}. In the following, we recall the main idea of the second class of approaches.

\begin{figure}[htbp]
  \centering
  \label{fig:forward-backward-diffusion}
  \includegraphics[width=\textwidth]{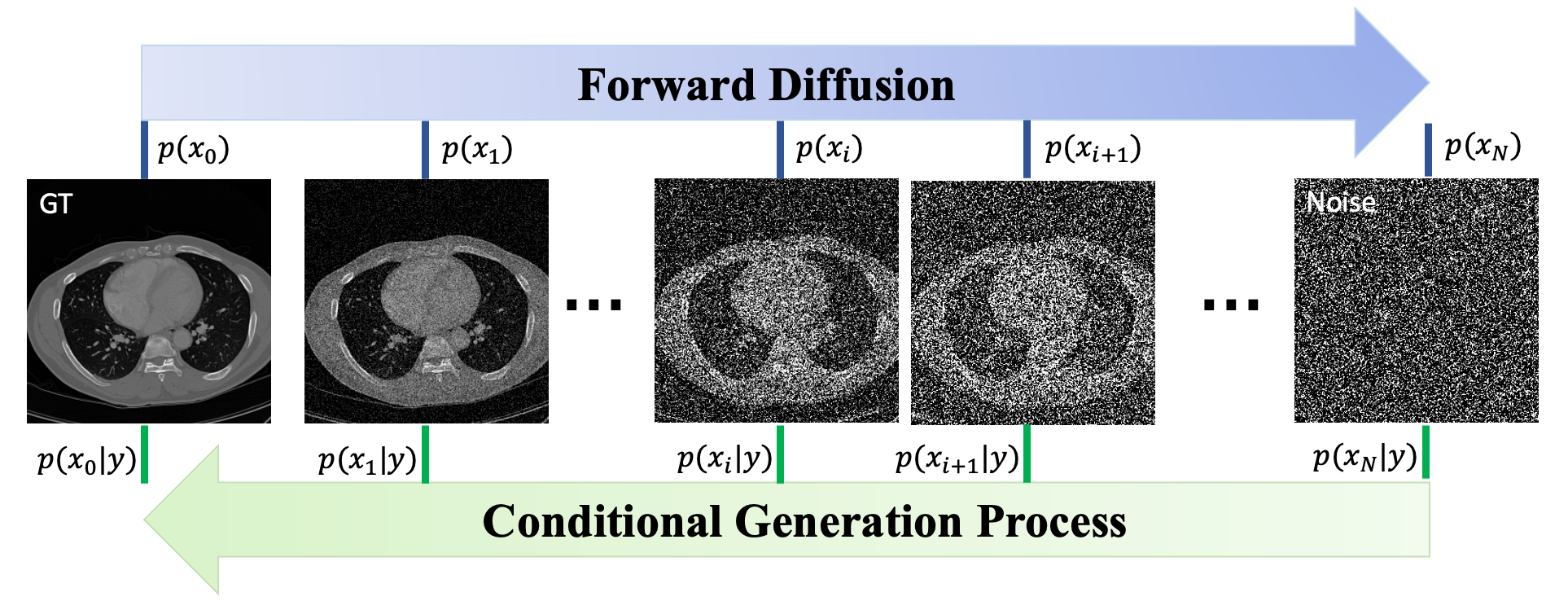}
  \caption{Forward diffusion and the conditional generation process.
  }
\end{figure}

The unconditioned diffusion model in \eqref{eq:reverse-diffusion-continuous-variable} (or \eqref{eq:DDPM-xk-to-xkm1}) provides us powerful data priors that use a sample $\bm{x}_T\sim p_{T}(\bm{x}_T)=\mathcal{N}(0, \bm{I})$ (a latent code) to generate a sample image $\bm{x}_0\sim p_0(\bm{x}_0)$. Therefore, the linear inverse problem in imaging (i.e., image inpainting, debluring, MRI, and CT) can be solved with a plug-in diffusion image prior. To incorporate the measurement $\bm{y}$, a posterior sampler is constructed by approximating the noisy sample distribution $p_t(\bm{x}_t)$ in the score function $\nabla_{\bm{x}_t}\log p_t(\bm{x}_t)$ by the posterior distribution $p(\bm{x}_t|\bm{y})$. \Cref{fig:forward-backward-diffusion} shows the general flow chart of the forward diffusion and the backward conditional generation process.

The reverse time diffusion process with a measurement constraint can be formulated by an SDE as
\begin{equation}\label{eq:reverse-diffusion-conditional}
    {\rm{d}}\bm{x}(t)=\left(f(\bm{x}_t,t)-g(t)^{2}\nabla_{\bm{x}_t}\log p_t(\bm{x}_t|\bm{y})\right){\rm{d}}t+g(t){\rm{d}}\bar{\bm{w}}.
\end{equation}
Note that this SDE does not directly correspond to the original forward diffusion model in \eqref{eq:forward-diffusion-SDE}. Even though the modification is straightforward, the exact posterior sampling for the diffusion model is usually intractable. 
Based on the Bayes's rule,  we have 
$$\nabla_{\bm{x}}\log p(\bm{x}|\bm{y})= \nabla_{\bm{x}}\log p(\bm{y}|\bm{x})+\nabla_{\bm{x}}\log p(\bm{x}).$$
Thus, we can deduce a equivalent form SDE as 
\begin{equation}\label{eq:reverse-diffusion-plug-in-diffusion-prior}
    {\rm{d}}\bm{x}(t)=\left[f(\bm{x},t)-g(t)^{2}(\nabla_{\bm{x}_t}\log p(\bm{y}|\bm{x}_t)+\nabla_{\bm{x}_t}\log p_t(\bm{x}_t))\right]{\rm{d}}t+g(t){\rm{d}}\bar{\bm{w}}.
\end{equation}
In this model, the score function $\nabla_{\bm{x}_t}\log p_t(\bm{x}_t)$ can be estimated by the pre-trained model $s_{\bm{\theta}^\ast}(\cdot,\cdot)$. However, the likelihood $p(y|\bm{x}_t)$ is another challenge in the computation because it does not have an analytic expression in the general inverse problem. To circumvent the challenge,  the likelihood  function  $p(\bm{y}|\bm{x}_t)$ is factorized as 
\begin{align}\label{eq:likelihood-func}
    p(\bm{y}|\bm{x}_t)=\int p(\bm{y}|\bm{x}_t,\bm{x}_0)p(\bm{x}_0|\bm{x}_t)d\bm{x}_0= \int p(\bm{y}|\bm{x}_0)p(\bm{x}_0|\bm{x}_t)d\bm{x}_0,
\end{align}
where the second equation uses the fact  that both $\bm{y}$ and $\bm{x}_t$ are conditionally independent on $\bm{x}_0$. In \cite{chung2023DPS}, this integration is approximated  by $p(\bm{y}|\hat{\bm{x}}_{0t})$ as 
$$p(\bm{y}|\bm{x}_t)\simeq p(\bm{y}|\hat{\bm{x}}_{0t})$$
where $\hat{\bm{x}}_{0t}$ represents the denoised data of the noisy sample $\bm{x}_t$. This is equivalent to that $p(\bm{x}_0|\bm{x}_t)$ is approximated by a delta distribution.
The approximation error is bounded by an upper bound that depends on the measurement error and the norm of the forward imaging operator \cite{chung2023DPS}. Authors in \cite{song2023LGD} improve the posterior estimation by the Monte Carlo approach where multiple samples are adopted to approximate the integration in \eqref{eq:likelihood-func}.

For the linear inverse problem $\bm{y}=\bm{Ax}+\bm{\epsilon}$ with $\bm{\epsilon}\sim \mathcal{N}(0,\sigma^2\bm{I})$, authors in \cite{jalal2021robust} proposed to approximate the score function of the posterior $p(\bm{y}|\bm{x}_t)$  by 
$$\nabla_{\bm{x}}\log p(\bm{y}|\bm{x})\simeq \frac{\bm{A}^{H}(\bm{y}-\bm{Ax})}{\sigma^2+\gamma_{t}^2},$$
with the hyperparameter sequence $\{\gamma_{t}\}_{t=1}^{N}$ are annealed during the generation process, $\bm{A}^{H}$ means the Hermitian transpose of forward imaging operator $\bm{A}$. This heuristic approach can only be used for linear inverse problems, and it is hard to handle the measurement noise in $\bm{y}$.

Even though the approximated posterior samplers do not have a theoretical guarantee to converge to the correct distribution in polynomial time as pointed out in \cite{gupta2024diffusion}, numerical results show quite plausible image processing and generation performance in various applications \cite{chung2023DPS,song2021scoreSDE,chung2022accMRI,chung2022come}. In this work, we propose a novel conditional generation framework with a diffusion image prior to reconstructing the MSCT volume data. More details are presented in Section \ref{sec:our-methods}. 

\subsection{ Multi-Source Static CT}\label{sec:msxs}
\begin{figure}[htbp]
  \centering
  \label{fig:cbct-system}\includegraphics[width=0.84\textwidth]{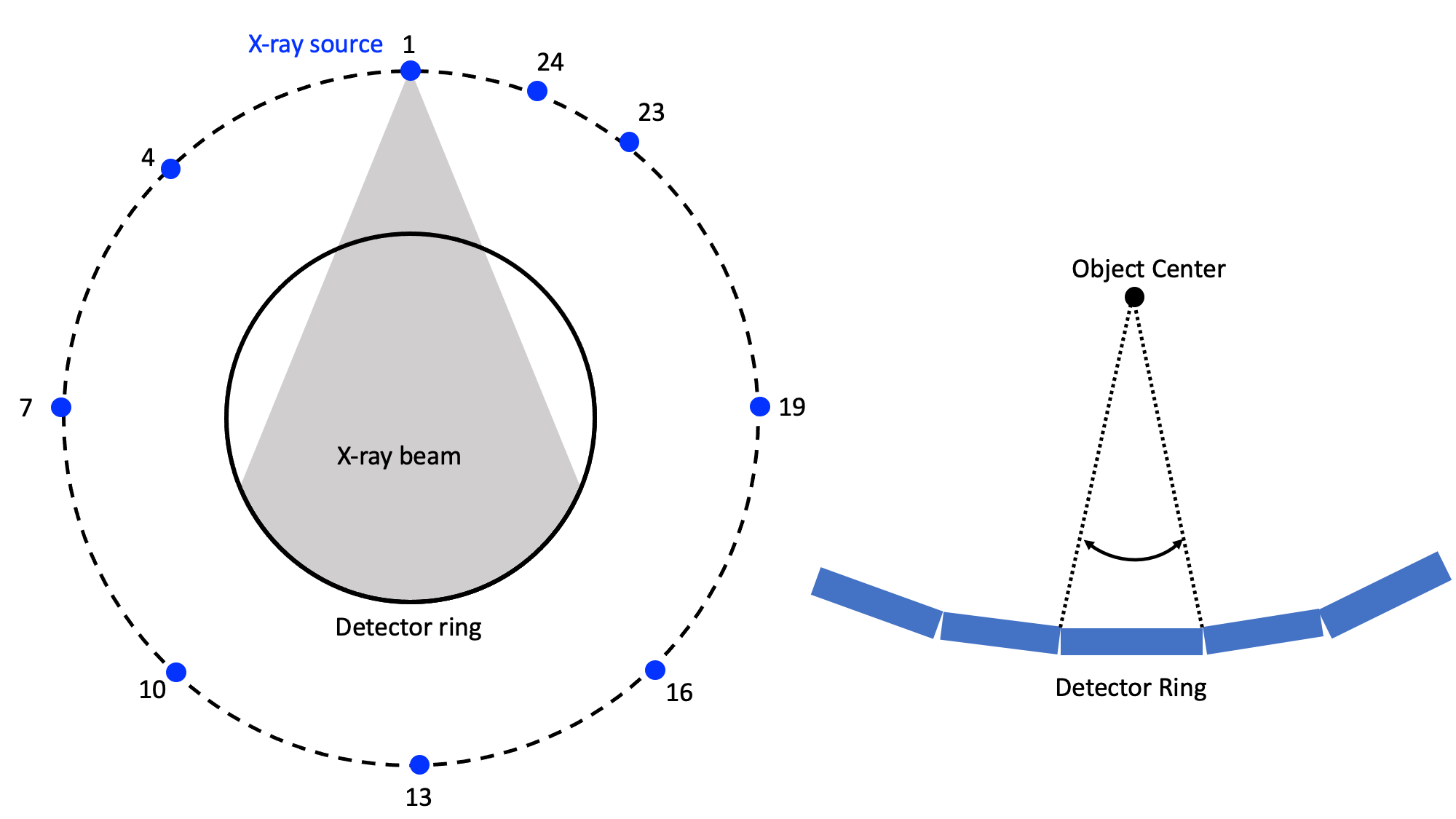}
  \caption{ Multi-Source Static CT System.
  }
\end{figure}

To accelerate the scanning speed of the CT system, Nanovision Technology (Beijing) Co., Ltd. has designed a CompoundEyeCT imaging system equipped with multiple static X-ray sources for CT. The system has 24 X-ray sources that are equally distributed around a ring covering 360 degrees. The detectors are fixed as a ring belt marked by a solid bold circle as shown in Figure \ref{fig:cbct-system}. The detector ring is composed of a regular 64-gon formed by 64 flat-panel detectors. A single flat-panel detector board has a fan angle $5.625^\circ$ as shown in the right of Figure \ref{fig:cbct-system}. For each X-ray source focus point, the covered detector arrays are in a cone beam shape. The measured projection of each view is a rectangle array (or 2D matrix). Since all sources can be controlled by a pulse signal, the Multi-Source Static CT imaging system can quickly acquire 24 view projections in a few milliseconds. Each time a small angle increment shifts the sources, we can obtain another group of 24 views' projection.  
The left sub-figure in Figure \ref{fig:cbct-system} shows the geometry of the CompoundEyeCT system. 
The blue dots on the dashed line circle show the fixed position of the X-ray sources.

\subsubsection{Scanning Mode}
\begin{figure}[htbp]
  \centering
  \label{fig:sparse-scan-traj}\includegraphics[width=0.4\textwidth]{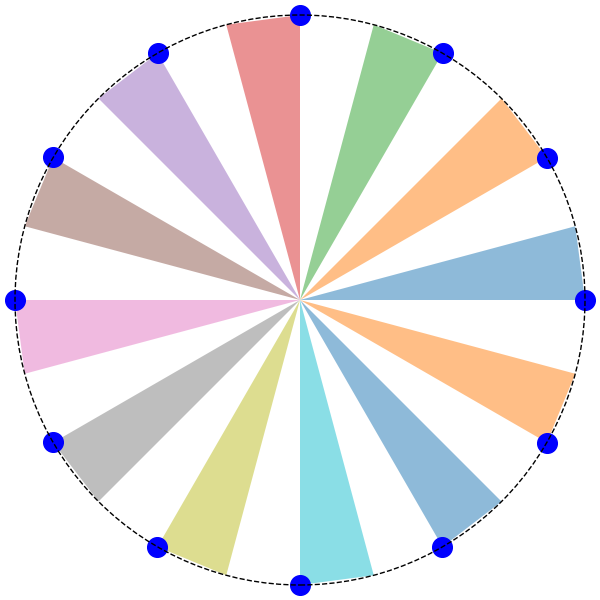}
  \caption{Sparse and non-uniform scanning trajectory. The small blue circle disk indicates the start position of the X-ray source.The arc edge of the white fan-shaped area are not scanned in the short scan mode.
  }
\end{figure}

Short scan settings in the CT imaging system can reduce the radiation dose and scanning time. The mechanical structure of the MSXS CompoundEyeCT equipment makes it a flexible system that controls the scanning direction/range and speed. Therefore, we can obtain a sparse view and limited angle projection with the scanning trajectory as shown in Figure \ref{fig:sparse-scan-traj}. The scanning views are sparsely distributed around the arc edge of the colorful fan-shaped area, and white fan regions are not covered during scanning. This scanning mode corresponds to a novel incomplete data Multi-Source Static CT volume reconstruction problem.

\section{Methods}\label{sec:our-methods}
This section introduces the diffusion image prior-driven MSCT volume reconstruction algorithm. Reverse time diffusion is adopted as a strong image prior to generating the 3D phantom from a latent encoding tensor. Then, a new diffusion posterior sampler is designed to generate the phantom with a measurement constraint. Finally, we summarize the whole MSCT volume reconstruction algorithm.

\subsection{Diffusion Image Prior for MSCT}\label{sec:dip-for-cbct}
To solve the 3DCT imaging problem in \eqref{eq:linear-system}, we incorporate measurement $\bm{Y}$ into the reverse time diffusion process. General approaches to sampling the diffusion posterior are reviewed in Subsection \ref{sec:conditional-generation-review}.
Assume that we have an unconditional diffusion model and a pre-trained neural network that approximated score function $s_{\bm{\theta}^\ast}$.
The state transition $p(\bm{x}_{t-1}|\bm{x}_{t})$ is modified as $p(\bm{x}_{t-1}|\bm{x}_{t}, \bm{Y})$ to explicitly reflect the data-dependent generation. The imaging physics is incorporated into the posterior sampling for task-driven 3DCT volume reconstruction.

\begin{figure}[htbp]
  \centering\label{fig:reverse-diffusion-process}
  \includegraphics[width=\textwidth]{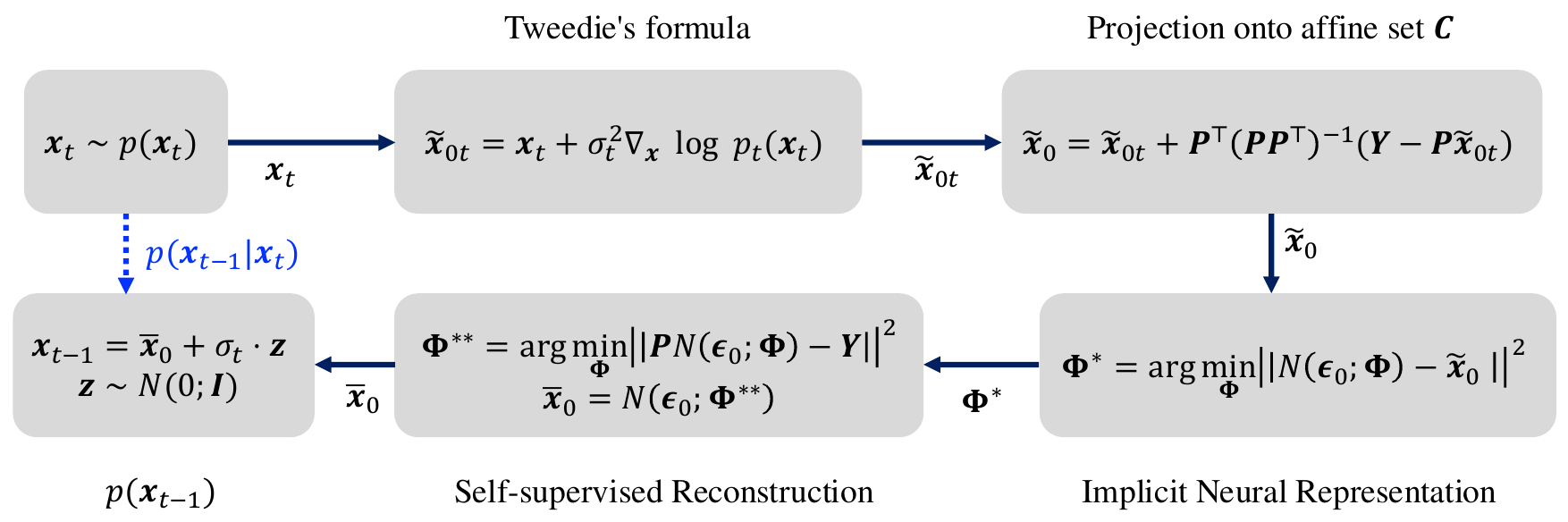}
  \caption{The flow chart of the diffusion posterior sampling scheme in the reverse time diffusion process $p(\bm{x}_{t-1}|\bm{x}_{t})$.
  }
\end{figure}

The main framework of our posterior sampling process is established as follows. 
For a currently sampled $\bm{x}_{t}\sim p_t(\bm{x}_t|\bm{x}_{t+1})$, it can be viewed as a noisy version of $\bm{x}_{t-1}$. So we first estimate a noiseless image $\tilde{\bm{x}}_{0t}$ that is assumed to be lay close to $p(\bm{x}_0)$.
Then, $\tilde{\bm{x}}_{0t}$ is projected onto the solution subspace $\bm{C}=\{\bm{x}|\bm{Px}=\bm{Y}\}$, and the projector is denoted as $\tilde{\bm{x}}_{0}$. We can obtain a continuous representation of the phantom in the spatial domain by adopting an implicit neural representation of the voxel data $\tilde{\bm{x}}_{0}$. To reduce the accumulated error in the former steps, we adopt the self-supervised learning algorithm to enhance the reconstructed image. Finally, we need to simulate a data point to return to the $t-1$ timestamp data distribution $p(\bm{x}_{t-1})$ of the reverse diffusion process. The simulated point $\bm{x}_{t-1}$ is obtained by adding a properly defined noise to the reconstructed image $\bar{\bm{x}}_{0}$ by the self-supervised learning algorithm. The whole framework is summarized in Figure \ref{fig:reverse-diffusion-process}. Details of the conditional generation steps between $p(\bm{x}_t)$ and $p(\bm{x}_{t-1})$ are explained in the following subsections. 

\subsubsection{Projection onto Affine Set}
Given an image sample $\bm{x}_t$ at timestamp $t$, the posterior expectation of the noiseless image can be computed by Tweedie's formula \cite{efron2011tweedie}
$$\mathbb{E}[\bm{x}_{0}|\bm{x}_t]=\bm{x}_{t}+\sigma^{2}_{t}\nabla_{\bm{x}}\log p_t(\bm{x}_t).$$
This expectation is essentially a minimum mean squared error (MMSE) estimator of the noiseless data $\bm{x}_0$ given the noisy sample $\bm{x}_t$ and the noise level $\sigma_t$. 
When the score function $\nabla_{\bm{x}}\log p_t(\bm{x}_t)$ is replaced by a pre-trained neural network approximated score function $s_{\bm{\theta}^\ast}(\cdot,\cdot)$, we have the denoised image  with the following form
\begin{equation}\label{eq:Tweedie-NN-score-x0t}
\tilde{\bm{x}}_{0t}=\tilde{\bm{x}}_{0t}(\bm{x}_t)=\bm{x}_t+\sigma_{t}^2 s_{\theta^\ast}(\bm{x}_t,t),
\end{equation}
where $\sigma_t$ is a pre-defined noise level-related parameter.
Since this noiseless estimation, $\tilde{\bm{x}}_{0t}$ is not directly dependent on the measurement $\bm{Y}$, it is projected onto the solution subspace $\bm{C}=\{\bm{x}|\bm{Px}=\bm{Y}\}$ and its projector is denoted by the new variable $\tilde{\bm{x}}_{0}$.

Now, for convenience of presentation, we introduce the definition of projecting a variable onto an affine set.
\begin{definition}\cite{bauschke2019convex}\label{def:affine-set-projector}
    Suppose that the range of linear operator $\bm{A}$ is closed and the operator $\bm{AA}^\ast$ is invertible. Define an affine set $\bm{C}=\{\bm{x}|\bm{Ax}=\bm{y}\}$, then for an arbitrary $\bm{x}$, the projector onto affine subspace $\bm{C}$ is defined by
\begin{equation}\label{eq:def-affine-set-projector}
    \bm{\mathbb{P}}_{\bm{C}}(\bm{x})=\bm{x}+\bm{A}^{\top}(\bm{AA}^{\top})^{-1}(\bm{y}-\bm{Ax}).
\end{equation}
\end{definition}

In the reverse time diffusion process, the generated $\tilde{\bm{x}}_{0t}$  is expected to be close to the measurement constraint set $\bm{C}=\{\bm{x}|\bm{Px}=\bm{Y}\}$. 
Based on Definition \ref{def:affine-set-projector}, the projector of $\tilde{\bm{x}}_{0t}$ is defined by
\begin{equation}\label{eq:x0t-projector}
    \tilde{\bm{x}}_0=\bm{\mathbb{P}}_{\bm{C}}(\tilde{\bm{x}}_{0t})=\tilde{\bm{x}}_{0t}+\bm{P}^{\top}(\bm{PP}^{\top})^{-1}(\bm{Y}-\bm{P}\tilde{\bm{x}}_{0t}).
\end{equation}

In 3DCT imaging, it is challenging to explicitly compute the inverse of the operator $\bm{PP}^{\top}$. Thus, we adopt the conjugate gradient (CG) algorithm to approximately compute the inverse operator $(\bm{PP}^{\top})^{-1}$. This is equivalent to solving the following optimization problem
$$\bm{y}^{\ast}_{CG}=\arg\min_{\bm{y}}||(\bm{PP}^{\top})\bm{y}-(\bm{Y}-\bm{P}\tilde{\bm{x}}_{0t})||^2.$$
Now, we obtain the projector of $\tilde{\bm{x}}_{0t}$ as
\begin{equation}\label{eq:x0t-projector-approx}
    \tilde{\bm{x}}_0=\bm{\mathbb{P}}_{\bm{C}}(\tilde{\bm{x}}_{0t})=\tilde{\bm{x}}_{0t}+\bm{P}^{\top}\bm{y}^{\ast}_{CG}.
\end{equation}
We should note that there will always be noise in the measured projection $\bm{Y}$ in practice. Therefore, this project onto an affine set operation is not optimal. However, it is quite simple to implement and is important to guide the conditional generation process. Thus, we will reduce the noise-caused error in the projection step in later substeps.

\subsubsection{INR for Phantom Representation}
The main idea of INR in Subsection \ref{sec:INR-background} shows that a neural network with a low-dimensional encoding tensor can represent a 3D phantom in continuous domain $[0,1]^3$. For example, we choose the neural network input tensor $\bm{\epsilon}_0\in\mathbb{R}^{M\times N \times K}$ with 
$$\bm{\epsilon}_0(i,j,k)=(i/M,j/N,k/K), i=0,1,...,M-1; j=0,1,...,N-1; k=0,1,...,K-1,$$
where $M,N,K$ represents the number of voxel bins on the $x,y,z$ axes. Then, we adopt a neural network $\mathcal{F}(\bm{\epsilon}_0; \bm{\Phi})$ with model parameters $\bm{\Phi}$ to represent the projector $\tilde{\bm{x}}_0$ in \eqref{eq:x0t-projector-approx}. 
To find the optimal parameters $\bm{\Phi}^\ast$ for the 3D phantom representation of $\tilde{\bm{x}}_0$, the optimization model can be written as 
\begin{equation}\label{eq:INR-model}
\bm{\Phi}^\ast=\arg\min_{\bm{\Phi}}\|\mathcal{F}(\bm{\epsilon}_0; \bm{\Phi})-\tilde{\bm{x}}_0\|^2+ R_{\lambda}(\bm{\Phi}).
\end{equation}
The objective function in this model is used as a loss function during the neural network training. The first term preserves the consistency of the data between the phantom represented by the neural network and the target phantom $\tilde{\bm{x}}_0$.
The second term $R_{\lambda}(\bm{\Phi})$ is a regularization term to stabilize neural network training \cite{loshchilov2019decoupled,hanson1988comparing} and $\lambda\in\mathbb{R}_{+}=\{\lambda| \lambda>0, \lambda\in\mathbb{R}\}$ is a hyper-parameter to balance the data fitting term and the regularization term of the parameters. 
Once the neural network $\mathcal{F}$ is well trained and the optimal parameters $\Phi^\ast$ are obtained, the 3D phantom can be continuously represented in the spatial domain. When we need a phantom with the required resolution, it can be obtained by resampling the unit cube $[0,1]^3$  with a larger number of discretization bins while the encoding tensor $\bm{\epsilon}_0$ is set to
$\bm{\epsilon}_{\mbox{test}}\in\mathbb{R}^{rM\times rN\times rK}$. Here, $r\ge 1$ is a positive integer $r\in\mathbb{N}_{+}$. Now, we obtain the predicted reconstruction (3D phantom) from the trained neural network as
\begin{equation}\label{eq:NeRF-hat-x0}
    \hat{\bm{x}}_0=\mathcal{F}(\bm{\epsilon}_{\mbox{test}}; \bm{\Phi}^\ast).
\end{equation}
This newly resampled phantom has a higher resolution ($r$ times) than the target phantom $\tilde{\bm{x}}_0$ during training. This design of any resolution reconstruction will be more attractive in practical applications.

\subsubsection{Self-supervised Learning (SSL) for MSCT}
Once we adopt the INR to represent $\tilde{\bm{x}}_0$ as in \eqref{eq:INR-model}, the reconstructed phantom \eqref{eq:NeRF-hat-x0} can be further enhanced by an SSL-based reconstruction algorithm to reduce the accumulation of errors caused by the reverse time diffusion process and the noise in the measured projection $\bm{Y}$. More precisely, the trained neural network $\mathcal{F}_{\bm{\Phi}^{\ast}}$ in \eqref{eq:INR-model} can be used as an initialization. Then, we adopt the following objective function to fine tune the neural network $\mathcal{F}_{\bm{\Phi}}$ and obtain an enhanced reconstructed 3D phantom
\begin{equation}\label{eq:ssl-model}
    \bm{\Phi}^{\ast\ast}=\arg\min_{\bm{\Phi}} \mathcal{L}(\bm{\Phi})=\|\bm{P}\mathcal{F}(\bm{\epsilon}_0; \bm{\Phi})-Y\|^2+ R_{\lambda}(\bm{\Phi}),
\end{equation}
where the second term $R_{\lambda}(\bm{\Phi})$ is chosen to stabilize the neural network training process as in \eqref{eq:INR-model}. On the one hand, the regularization term $R_{\lambda}(\bm{\Phi})$ can be chosen as a weight penalty based on the $L_2$ norm. On the other hand, it can be further extended as a total variation (TV)  norm \cite{rudin1992ROF} as  follows
$$R_{\lambda}(\bm{\Phi})=\lambda\|\nabla \mathcal{F}(\bm{\epsilon}_0; \bm{\Phi})\|_2^2$$
to penalize the smoothness of the reconstructed 3D phantom with a hyper-parameter $\lambda\in \mathbb{R}_{+}$. 
In this objective function, only the projection $\bm{Y}$ is used to supervise the reconstruction of the MSCT volume data. Thus, this model is preferred in practice because the ground truth phantom is scarce in the supervised learning based deep learning models. The finally enhanced reconstructed phantom is obtained from the well-trained neural network $\mathcal{F}_{\bm{\Phi}}$ and denoted by 
\begin{equation}\label{eq:NAF-phant-x0-bar}
    \bar{\bm{x}}_0=\mathcal{F}(\bm{\epsilon}_0; \bm{\Phi}^{\ast\ast}).
\end{equation}
Here, the input tensor $\bm{\epsilon}_{0}$ can be chosen as explained in \eqref{eq:NeRF-hat-x0} with a predefined target reconstruction resolution.

The above INR and SSL-based reconstruction model can be summarized to get a joint optimization model as
$$\bm{\Phi}^{\ast\ast}=\arg\min_{\bm{\Phi}}\|\bm{P}\mathcal{F}(\bm{\epsilon}_0; \bm{\Phi})-\bm{Y}\|^2+\mu\|\mathcal{F}(\bm{\epsilon}_0; \bm{\Phi})-\tilde{\bm{x}}_0\|^2+\lambda R(\bm{\Phi}),$$
where $\mu>0$ is a hyper-parameter to balance the INR and SSL term. If the INR is replaced by the original variable $\bm{x}$ and chooses the TV-norm regularization term, this model returns to the classical image reconstruction model
\begin{equation}
\bar{\bm{x}}_{0}=\arg\min_{\bm{x}}\|\bm{Px}-\bm{Y}\|^2+\mu \|\bm{x}-\tilde{\bm{x}}_0\|^2+\lambda\|\nabla \bm{x}\|_{1}.
\end{equation}
This model indicates that, on the one hand, the trained neural network $\mathcal{F}_{\bm{\Phi}}(\cdot;\cdot)$ fits the projector $\tilde{\bm{x}}_0$ in the affine subspace $\bm{C}$. On the other hand, the SSL framework refines the reconstructed phantom to satisfy the physical imaging model \eqref{eq:linear-system}.

\subsubsection{Pseudo-forward State Transition}
Based on the forward diffusion model, we can obtain the image sample $\bm{x}_{t-1}\sim p_{t-1}(\bm{x}_{t-1})$ at timestamp $(t-1)$ as 
$$\bm{x}_{t-1}=\bar{\bm{x}}_0+\sigma_{t-1}\cdot \bm{z}, \quad \bm{z} \sim \mathcal{N}(0, \bm{I})$$
where $\bm{z}$ is the white Gaussian noise with the same size as the reconstructed phantom $\bar{\bm{x}}_0$ in \eqref{eq:NAF-phant-x0-bar}. The noise level is set based on the chosen diffusion scheme. In summary, we finish the state transition from $p_t(\bm{x}_t)$ to $p_{t-1}(\bm{x}_{t-1})$ with the substeps \eqref{eq:Tweedie-NN-score-x0t}, \eqref{eq:x0t-projector}, and \eqref{eq:INR-model}-\eqref{eq:NAF-phant-x0-bar} which constitute a conditional generation step.

\subsection{Algorithm Summarization}\label{sec:alg-summary}
When the state is transited from $\bm{x}_{T}$ to $\bm{x}_0$ as described in Subsection \ref{sec:dip-for-cbct}, we obtain the reconstructed 3D phantom from the MSCT scanning data $\bm{Y}$. The proposed algorithm combines the deep diffusion image prior, affine set projection, the INR-based phantom representation, and the SSL-based phantom reconstruction; we denote it as DIP-ASPINS. The pseudo-code of the proposed DIP-ASPINS algorithm for MSCT imaging is summarized in Algorithm \ref{alg:DIP-ASPINS}.

\begin{algorithm}
\caption{DIP-ASPINS for MSCT
}
\label{alg:DIP-ASPINS}
\begin{algorithmic}
\STATE Input: projection $\bm{Y}$,
SDE discretization steps $N$, conditional diffusion update interval $K$, pre-trained score function $s_{\bm{\theta}^\ast}$, the random noise tensor $\bm{x}_{N}\sim \mathcal{N}(0,\bm{I})$
\STATE Initialization: $\bm{\sigma}_{t},t=1,...,N$
\STATE \# Reverse time conditional diffusion 
\FOR{$t=N:1$}
\IF{$t\%K==0$}
\STATE \# Conditional generation
\STATE Tweedie's formula for denoising
$\tilde{\bm{x}}_{0t}=\bm{x}_{t}+\bm{\sigma}^{2}_{t}s_{\bm{\theta}^\ast}(\bm{x}_t,t)$
\STATE Affine set projection $\tilde{\bm{x}}_0=
\tilde{\bm{x}}_{0t}+\bm{P}^{\top}(\bm{PP}^{\top})^{-1}(\bm{Y}-\bm{P}\tilde{\bm{x}}_{0t})$
\STATE INR for 3D phantom representation:\\
\qquad (1)$
\bm{\Phi}^{\ast}=\mathop{\arg\min}\limits_{\bm{\Phi}}\|\mathcal{F}(\bm{\epsilon}_0,\bm{\Phi})-\tilde{\bm{x}}_0\|^2+R_{\lambda}(\bm{\Phi}), \quad \bm{\epsilon}_{0} \sim \mathcal{N}(0,\bm{I}),$\\
\qquad (2) $\hat{\bm{x}}_0=\mathcal{F}(\bm{\epsilon}_{00};\bm{\Phi}^\ast), \quad \bm{\epsilon}_{00} \sim \mathcal{N}(0,\bm{I}),$
\STATE SSL for MSCT reconstruction:\\
\qquad (1)
$\bm{\Phi}^{\ast\ast}=\mathop{\arg\min}\limits_{\bm{\Phi}}\|\bm{P}\mathcal{F}(\bm{\epsilon}_0,\bm{\Phi})-Y\|^2+R_{\lambda}(\bm{\Phi}), \quad \bm{\epsilon}_{0} \sim \mathcal{N}(0,\bm{I}),$\\
\qquad (2) $\bar{\bm{x}}_{0}=\mathcal{F}(\bm{\epsilon}_{01};\bm{\Phi}^{\ast\ast}), \quad \bm{\epsilon}_{01} \sim \mathcal{N}(0,\bm{I}),$
\STATE Next image prior
$\bm{x}_{t-1}=\bar{\bm{x}}_0+\bm{\sigma}_{t-1}\cdot \bm{z}, \quad \bm{z}\sim \mathcal{N}(0,\bm{I})$
\ELSE
\STATE \# Unconditional generation
\STATE Generation process: $\displaystyle \bm{x}_{t-1}=\bm{x}_{t}+(\bm{\sigma}_{t}^{2}-\bm{\sigma}_{t-1}^{2})s_{\bm{\theta}^{\ast}}(\bm{x}_t,t)+\sqrt{\bm{\sigma}_{t}^{2}-\bm{\sigma}_{t-1}^{2}}\bm{\epsilon}, \bm{\epsilon} \sim \mathcal{N}(0,\bm{I})$
\ENDIF
\ENDFOR
\RETURN Output: $\bm{x}_0.$
\end{algorithmic}
\end{algorithm}

If we remove the INR and SSL module in DIP-ASPIN Algorithm \ref{alg:DIP-ASPINS}, we obtain another simplified diffusion image prior based conditional generation model for MSCT reconstruction. This new algorithm is summarized in Algorithm \ref{alg:DIP-ASP} and is denoted by DIP-ASP. In this algorithm, the conditional generation process is guided by the affine set projection operation to incorporate the measured projection $\bm{Y}$ and to preserve the data consistency. Since this algorithm does not contain a learning process, it is more efficient than the DIP-ASPINS algorithm in the implementation. However, we should note that the affine set projection operation will induce an error in the reconstructed phantom whenever the measured projection $\bm{Y}$ is degraded by noise.

\begin{algorithm}
\caption{DIP-ASP for MSCT
}
\label{alg:DIP-ASP}
\begin{algorithmic}
\STATE Input: projection $\bm{Y}$, 
SDE discretization steps $N$, conditional diffusion update interval $K$, pre-trained score function $s_{\bm{\theta}^\ast}$, the random noise tensor $\bm{x}_{N}\sim \mathcal{N}(0, \bm{I})$
\STATE Initialization: $\bm{\sigma}_{t},t=1,...,N$
\STATE \# Reverse time conditional diffusion 
\FOR{$t=N:1$}
\IF{$t\%K==0$}
\STATE \# Conditional generation
\STATE Tweedie's formula for denoising
$\tilde{\bm{x}}_{0t}=\bm{x}_{t}+\bm{\sigma}^{2}_{t}s_{\bm{\theta}^\ast}(\bm{x}_t,t)$
\STATE Affine set projection $\tilde{\bm{x}}_0=
\tilde{\bm{x}}_{0t}+\bm{P}^{\top}(\bm{PP}^{\top})^{-1}(\bm{Y}-\bm{P}\tilde{\bm{x}}_{0t})$
\STATE Next image prior
$\bm{x}_{t-1}=\bar{\bm{x}}_0+\bm{\sigma}_{t-1}\cdot \bm{z}, \quad \bm{z}\sim \mathcal{N}(0, \bm{I})$
\ELSE
\STATE \# Unconditional generation
\STATE 
Generation process:
$\displaystyle \bm{x}_{t-1}=\bm{x}_{t}+(\bm{\sigma}_{t}^{2}-\bm{\sigma}_{t-1}^{2})s_{\bm{\theta}^{\ast}}(\bm{x}_t,t)+\sqrt{\bm{\sigma}_{t}^{2}-\bm{\sigma}_{t-1}^{2}}\bm{\epsilon}, \quad \bm{\epsilon} \sim \mathcal{N}(0,\bm{I})$
\ENDIF
\ENDFOR
\RETURN Output: $\bm{x}_0.$
\end{algorithmic}
\end{algorithm}

\section{Experimental results}\label{sec:experiments}
In this section, we test the performance of the proposed DIP-ASPINS Algorithm \ref{alg:DIP-ASPINS} and its variant DIP-ASP Algorithm \ref{alg:DIP-ASP} on MSCT (introduced in Subsection \ref{sec:msxs}) volume reconstruction tasks. The computing hardware is equipped with the NVIDIA RTX A6000 GPU (48G). The Adam optimizer \cite{kingma2015adam} with a learning rate $1\times 10^{-4}$ was adopted in the self-supervised learning algorithms of our models. The score function used in our reverse time diffusion process is a pre-trained model of the VE-SDE, which is introduced in \eqref{eq:VE-SDE}. The pre-trained score function model uses the same architecture as in the article \cite{song2021scoreSDE} and is trained on the AAPM dataset with data augmentation (random flipping and pixel value scaling). In the pre-training phase, we use Adam optimizer and set the gradient clipping norm to 1.0. For the learning rate scheduling, we first increase it linearly from 0 to $2\times10^{-4}$ during the first 5K steps. Then we keep a constant learning rate of $2\times10^{-4}$. This score function is trained by totally 1.5M steps.

\subsection{Comparison Methods} 
Due to the scarcity of a large-scale 3DCT dataset for deep supervised learning, we only compare the proposed methods to classical iterative reconstruction algorithms and the SSL-based methods. The iterative reconstruction approaches compared are (1) the conjugate gradient (CG) algorithm for the L2 model in \eqref{eq:L2+Ru} (denoted as L2-CG), and (2) the alternating direction method of multipliers (ADMM) algorithm for L2TV model \eqref{eq:L2+Ru} (denoted as L2TV-ADMM). The compared SSL-based methods are the recently published approaches named neural attention fields (NAF) \cite{zha2022naf} and diffusion prior driven neural representation (DPER) \cite{du2024dper}.

\subsection{INR Architecture}
The neural network for phantom representation is chosen to be a multilayer perceptron (MLP) with the hash encoding-based position embedding \cite{muller2022instantNGP}. The number of learnable parameters is $14.24 \mbox{M}$. 
The Adam optimizer is adopted to train the neural network in the 3D phantom implicit neural representation stage and the SSL volume reconstruction stage of our proposed Algorithm \ref{alg:DIP-ASPINS} and Algorithm \ref{alg:DIP-ASP}.

\subsection{ Data Preparation}
The test data is simulated by the phantoms from the ``2016 NIH-AAPM-Mayo Clinic Low Dose CT Grand Challenge'' data set (Abdomen) and the publicly accessed 3DCT imaging phantoms Pancreas and Stented Abdominal Aorta (SAA) \footnote{Pancreas and Stented Abdominal Aorta phantoms are downloaded from \href{https://klacansky.com/open-scivis-datasets/category-ct.html}{https://klacansky.com/open-scivis-datasets/category-ct.html}.}. All these phantoms are used as the ground truth for reconstruction algorithms' performance evaluation. The resolution of the Pancreas phantom is $512\times 512 \times 240$. The resolution of the SAA phantom is $512\times 512 \times 174$. For the AAPM dataset, we chose a phantom and rebin it along the z-axis to a thickness of 2mm per slice. Its spatial resolution is $512\times 512 \times 194$. 
We simulate the projection by forward projecting the phantoms using the MSCT system.
The noisy projection is simulated by adding Poisson noise and white Gaussian noise to the forward projection $Pu$ with the following formula
\begin{equation}\label{eq:add-noise-formula}
\bm{Y}=-\ln (\mbox{Poisson}(e^{-\bm{P}\bm{u}*I_0})/I_0)+\eta \cdot \bm{\epsilon}, \quad \bm{\epsilon} \sim \mathcal{N}(0,\eta^{2} \bm{I}),
\end{equation}
where $I_0$ is the X-ray source emitted photon intensity, $\bm{P}$ is the forward projection operator, and $\bm{u}$ is the 3D phantom. $\mbox{Poisson}(\cdot)$ denotes the simulation of the Poisson noise process. The higher value of $I_0$ corresponds to the lower-level Poisson noise. $\eta>0$ is the Gaussian noise level and $\bm{\epsilon}$ is the white Gaussian noise with the same shape as $\bm{Pu}$. $\bm{I}$ is the covariance matrix with diagonal values 1 and the else position 0 when the noise $\bm{\epsilon}$ and the data tensor are vectorized. 

For the proposed MSCT imaging system, the X-ray sources are rotated simultaneously during scanning with a predefined angle increment step ($\Delta \mbox{Angle})$. For example, when the number of sparse views is set to 120, each of the 24 X-ray sources rotates in the counterclockwise direction simultaneously for 5 steps, and each step is a $1^\circ$ rotation around the object center $O$.
Therefore, the X-ray sources leave a non-uniformly distributed trajectory as shown in \Cref{fig:sparse-scan-traj}.

\subsection{Evaluation of the DIP-ASPINS}
In this subsection, we will evaluate the proposed algorithms DIP-ASPINS (in Algorithm \ref{alg:DIP-ASPINS}) and the DIP-ASP (in Algorithm \ref{alg:DIP-ASP}) on the simulated noiseless and noisy data. The compared algorithms are tested on different numbers of sparse views, different noise levels, and phantoms.

\subsubsection{Noiseless Projection Reconstruction}
In this experiment, we simulate the projection of 3D phantoms without being degraded by noise. This setting is related to the fact that the affine set projection operation in the proposed DIP-ASPINS and DIP-ASP algorithms were designed with the noiseless constraint. We will test the performance of our proposed algorithms in a noisy projection setting in the next group of numerical experiments. In this noiseless volume data reconstruction study, the simulated sparse view projection is set to $\#\mbox{views}=48, 72, 120, 240$ with non-uniform distributed source trajectory as shown in \Cref{fig:sparse-scan-traj}. 
The SDE discretization steps in DIP-ASP and DIP-ASPINS are both set to $N=2000.$ The conditional generation process is updated at an interval $K=25$ when $t\in [1000,2000]$ and $K=50$ when $t\in[0,1000)$. The INR and SSL update steps are set to $N_{\mbox{INR}}=10$ and $N_{\mbox{SSL}}=50$ respectively. The compared methods, i.e., CG, ADMM, NAF, and DPER, are manually tuned to the optimal performance on the test data. 

The quantitative evaluation of the compared methods on sparse view MSCT reconstruction task is shown in \Cref{tab:quantative-results-3-phantoms-NonUniform-Scan}. The quality of the reconstructed volumes is measured by PSNR and SSIM \cite{psnr-ssim-2010,wang2004ssim}. We can observe that the proposed DIP-ASP and the DIP-ASPINS algorithms have better performance than the compared methods, i.e., L2-CG, L2TV-ADMM, NAF, and DPER when $\#\mbox{views}=48$ and $72$. DIP-ASPINS has the best SSIM values among the compared methods when the number of sparse views is increased to $120$ and $240$. However, the PSNR is inferior to NAF and DPER. These results indicate that the proposed DIP-ASPINS algorithm can produce 3D phantoms with better structure similarity to the ground truth than compared methods except the case $\# \mbox{views}=48$. When the sparse view case is set to $\# \mbox{views}=48$, the PSNR values of DIP-ASPINS are the best among compared methods on the phantoms Abdomen and Pancreas, and the DIP-ASP algorithm has the best PSNR and SSIM on SAA phantom. 

\begin{table}[htbp]
\footnotesize
\caption{Reconstruction results from non-uniformly distributed sparse view MSCT projection without noise. The test phantoms are Abdomen, Pancreas, and SAA. The number of views is set to $\#\mbox{views}=48, 72, 120, 240$.  
}\label{tab:quantative-results-3-phantoms-NonUniform-Scan}
\begin{center}
  \begin{tabular}{|c|c|c|c|c|c|} \hline
  
\multirow{2}{*}{ Phantom } & \multirow{2}{*}{ Methods } & \multicolumn{1}{c|}{\#views=48} & \multicolumn{1}{c|}{\#views=72} & \multicolumn{1}{c|}{\#views=120} & \multicolumn{1}{c|}{\#views=240} \\
\cline{3-6}
& &
PSNR/SSIM &  PSNR/SSIM &  PSNR/SSIM & PSNR/SSIM \\ \cline{1-6}

\multirow{6}{*}{ Abdomen } 
&L2-CG     & 23.41/0.5674 & 24.60/0.6631 & 25.55/0.7587 & 26.20/0.8345 \\ \cline{2-6} 
&L2TV-ADMM & 23.95/0.6216 & 24.80/0.6941 & 25.59/0.7783 & 26.11/0.8396  \\ \cline{2-6} 
&NAF       & 24.82/0.6260 & 25.57/0.6591 & \textbf{26.43}/0.7418 & 27.09/0.8460  \\ \cline{2-6} 
&DPER      & 23.84/0.7465 & 24.55/0.8056 & 25.04/0.8208 & 26.28/0.8691 \\ \cline{2-6} 
&DIP-ASP   & 24.42/\textbf{0.7980} & \textbf{26.34}/0.8220 & 26.40/0.8383 & 26.38/0.8368 \\ \cline{2-6} 
&DIP-ASPINS& \textbf{25.21}/0.7924 & 25.58/\textbf{0.8226} & 26.18/\textbf{0.8669} & \textbf{27.11}/\textbf{0.8801} \\ \cline{1-6} 

\multirow{6}{*}{\makecell[c]{Pancreas} } 
& L2-CG    & 23.67/0.6061 & 24.90/0.7140 & 25.72/0.7968 & 26.09/0.8371 \\ \cline{2-6} 
& L2TV-ADMM& 24.38/0.6890 & 25.12/0.7569 & 25.77/0.8232 & 26.12/0.8560 \\ \cline{2-6} 
& NAF      & 25.74/0.7288 & 26.49/0.7793 & 27.48/0.8668 & 27.84/0.9141 \\ \cline{2-6} 
& DPER     & 25.99/0.7951 & 26.01/0.8417 & \textbf{27.86}/0.8898 & \textbf{28.18}/0.9243 \\ \cline{2-6} 
& DIP-ASP  & 25.00/0.7808 & 25.33/0.7975 & 25.41/0.7966 & 26.54/0.8653  \\ \cline{2-6} 
& DIP-ASPINS& \textbf{26.46}/\textbf{0.8532} & \textbf{27.36}/\textbf{0.9090} & 27.47/\textbf{0.9149} & 28.07/\textbf{0.9376} \\ \cline{1-6} 

\multirow{6}{*}{ SAA } 
&L2-CG     & 26.50/0.6523 & 28.09/0.7463 & 29.53/0.8212 & 30.47/0.8664 \\ \cline{2-6} 
&L2TV-ADMM & 27.19/0.7179 & 28.30/0.7807 & 29.45/0.8422 & 30.23/0.8795 \\ \cline{2-6} 
&NAF       & 28.55/0.7955 & 29.80/0.8407 & 31.43/0.8866 & 32.64/0.9142 \\ \cline{2-6} 
&DPER      & 27.05/0.8304 & 27.25/0.8308 & 32.46/0.9217 & \textbf{33.59}/0.9314  \\ \cline{2-6} 
&DIP-ASP   & \textbf{31.74}/\textbf{0.8899} & 29.46/0.8617 & 29.79/0.8656 & 32.21/0.9007 \\ \cline{2-6} 
&DIP-ASPINS& 30.14/0.8762 & \textbf{31.68}/\textbf{0.8966} & \textbf{33.33}/\textbf{0.9281} & 33.49/\textbf{0.9338} \\ \cline{1-6} 
  \end{tabular}
\end{center}
\end{table}

\begin{figure*}[htbp]
  \centering
  \begin{tabular}{@{}p{0.98\linewidth}}
    \subfigimg[width=0.96\linewidth]{}{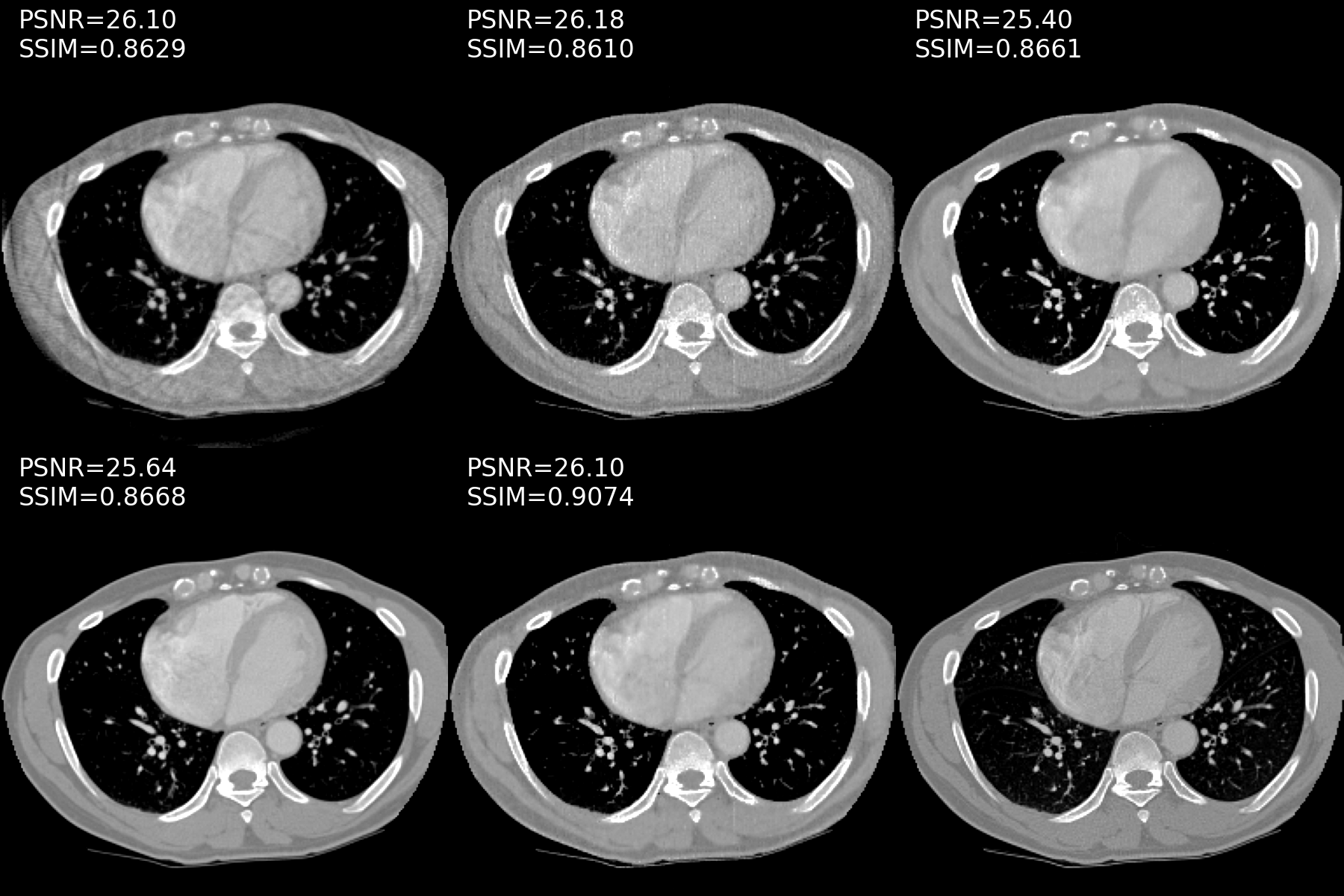} 
  \end{tabular}
\caption{The reconstructed transverse plane image slices of Abdomen phantom. The numbers of sparse views is $240$. From left to right, the first row shows the compared methods: ADMM, NAF, and  DPER. The second row shows the compared methods: DIP-ASP, DIP-ASPINS, and the ground truth. The display window is $[0.1, 0.6]$.
   }
\label{fig:rec-slices-BE006-65}
\end{figure*}

We choose the reconstructed transverse plane image slices from the Abdomen phantom and show them in \Cref{fig:rec-slices-BE006-65}. The number of projection views is set to 240.
The compared methods are L2TV-ADMM, NAF, DPER, DIP-ASP, DIP-ASPINS, and ground truth (GT). The L2-CG model's reconstruction result is similar but worse than L2TV-ADMM, so we omit showing this slice. 
From left to right, the first row of \Cref{fig:rec-slices-BE006-65} shows the reconstruction results from L2TV-ADMM, NAF, and DPER. The second row of \Cref{fig:rec-slices-BE006-65} shows the image slices of DIP-ASP, DIP-ASPINS, and GT. The visualization results show streak artifacts in the L2TV-ADMM algorithm's reconstruction. DPER produces images with higher SSIM value than NAF. The untrained DIP-ASP model produces an image slice with a better visualization effect than the NAF and DPER methods. The image slice from the proposed DIP-ASPINS has the best SSIM value among the compared methods. However, the PSNR value of DIP-ASPINS is lower than that of NAF, with a small gap. The NAF reconstruction has more noise artifacts in the center of the reconstructed images than the proposed DIP-ASPINS model. 

\begin{figure*}[htbp]
  \centering
  \begin{tabular}{@{}p{0.32\linewidth}@{}p{0.32\linewidth}@{}p{0.32\linewidth}}
    \subfigimg[width=\linewidth]{}{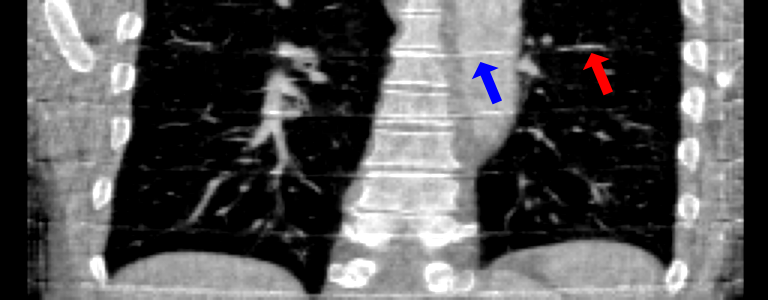} &
    \subfigimg[width=\linewidth]{}{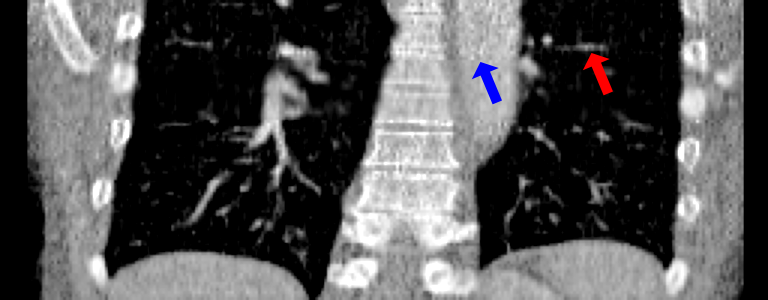} &
    \subfigimg[width=\linewidth]{}{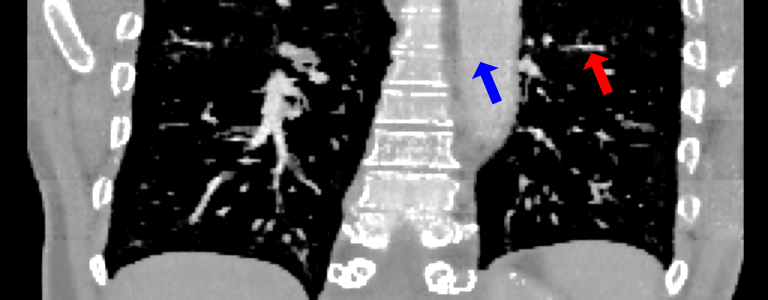} \\
    \subfigimg[width=\linewidth]{}{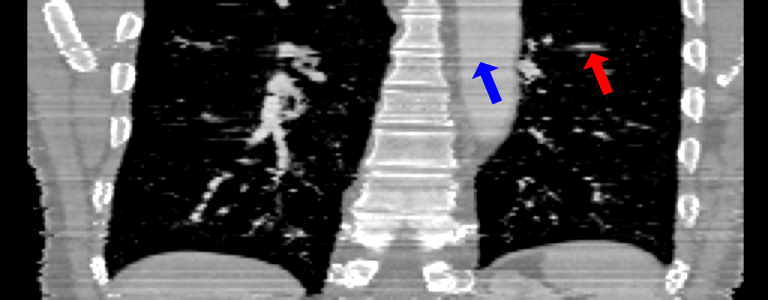} &
    \subfigimg[width=\linewidth]{}{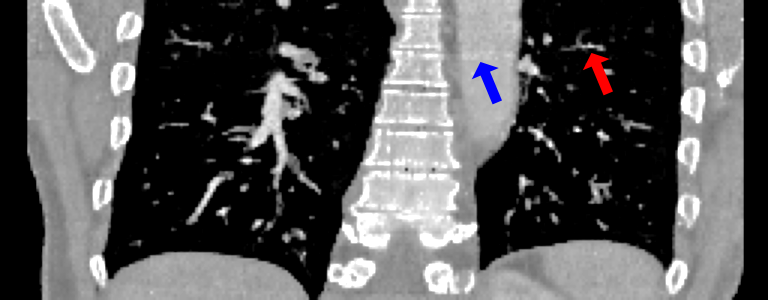} &
    \subfigimg[width=\linewidth]{}{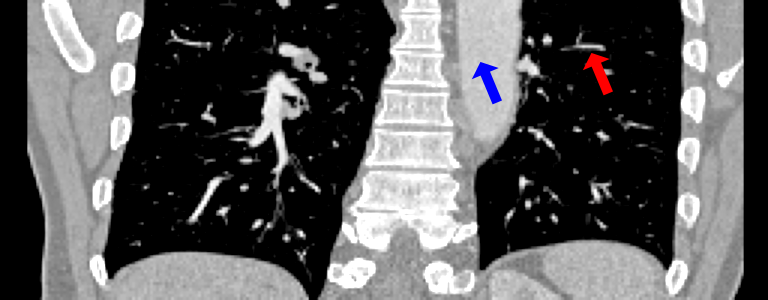}
  \end{tabular}
  \caption{The visualization of coronal plane image slices of Abdomen phantom. From left to right, the first row shows the compared methods: L2TV-ADMM, NAF, and  DPER. The second row shows the compared methods: DIP-ASP, DIP-ASPINS, and the ground truth. The display window is $[0.1, 0.6]$.
   }
\label{fig:rec-zy-slices-BE006-slice-174-views-120}
\end{figure*}

To visualize the consistency of the coronal plane image slice of the reconstructed Abdomen phantoms, we visualize the 174th slice from the compared methods in \Cref{fig:rec-zy-slices-BE006-slice-174-views-120}. The number of scanning views is set to 120.
The first row of \Cref{fig:rec-zy-slices-BE006-slice-174-views-120} shows the reconstructed image slices by the compared methods: ADMM, NAF, and DPER. The second row shows the image slices from DIP-ASP, DIP-ASPINS, and the ground truth. It is observed that the L2TV-ADMM and DIP-ASP methods in the first column contains streak artifacts (marked by blue arrow). The proposed DIP-ASPINS and the compared methods, DPER and NAF, show a smooth region around the blue arrow. However, there are noise artifacts around the blue arrow in the NAF reconstruction. For the bronchioles in the reconstructed slices (marked by red arrow), DIP-ASPINS and DPER show better structure similarity to ground truth than the compared methods. Other compared methods (ADMM, NAF, and DIP-ASP) produce image slices with inconsistent structure to the ground truth around the red arrow.

\subsubsection{Noisy Projection reconstruction}
\begin{figure}[htbp]
  \centering
  \includegraphics[scale=0.4235]{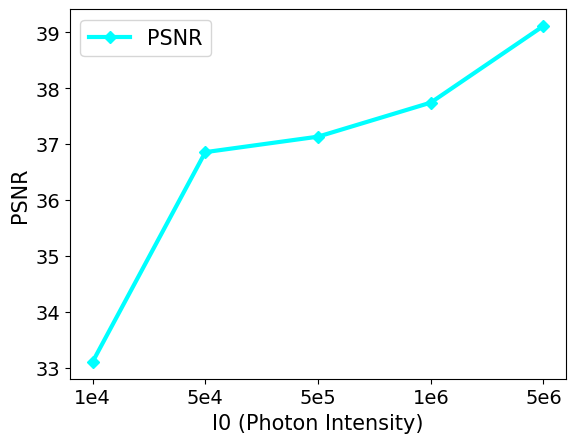}
  \hspace{6mm}
  \includegraphics[scale=0.4235]{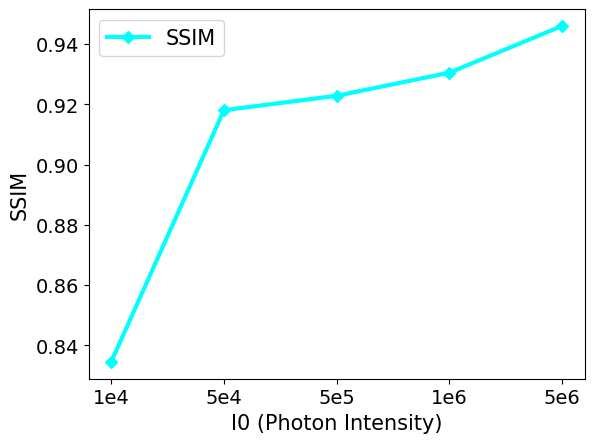}
  \caption{The PSNR (left subfigure) and SSIM (right subfigure) curves with respect to different noise levels. The test phantom is SAA with 20 slices, and the number of projection views is 120.}
  \label{fig:noisy-psnr-ssim}
\end{figure}

In this experiment, we set the noise level of the simulated projection in \eqref{eq:add-noise-formula} to $I_0=10^4, 5\times 10^4, 5\times 10^5, 10^6, 5\times 10^6.$ The Gaussian noise level is set to $\eta=0.05$. A phantom containing twenty consecutive slices of the SAA phantom is used as a test phantom. The number of sparse views is set to \#views=120. The PSNR and SSIM curves with respect to the different noise levels are shown in  \Cref{fig:noisy-psnr-ssim}. The number of steps in the generation process of the DIP-ASPINS model is set to $N=2000.$ The conditional generation update interval is set to $K=50$ at the first 500 steps and $K=25$ at the following 1500 diffusion steps. The results show that the quantitative measure of the reconstructed phantom by the DIP-ASPINS model will be improved when the noise level is lower (corresponding to larger values of $I_0$). 

\begin{figure*}[htbp]
  \centering
  \begin{tabular}{@{}p{0.98\linewidth}}
    \subfigimg[width=0.96\linewidth]{}{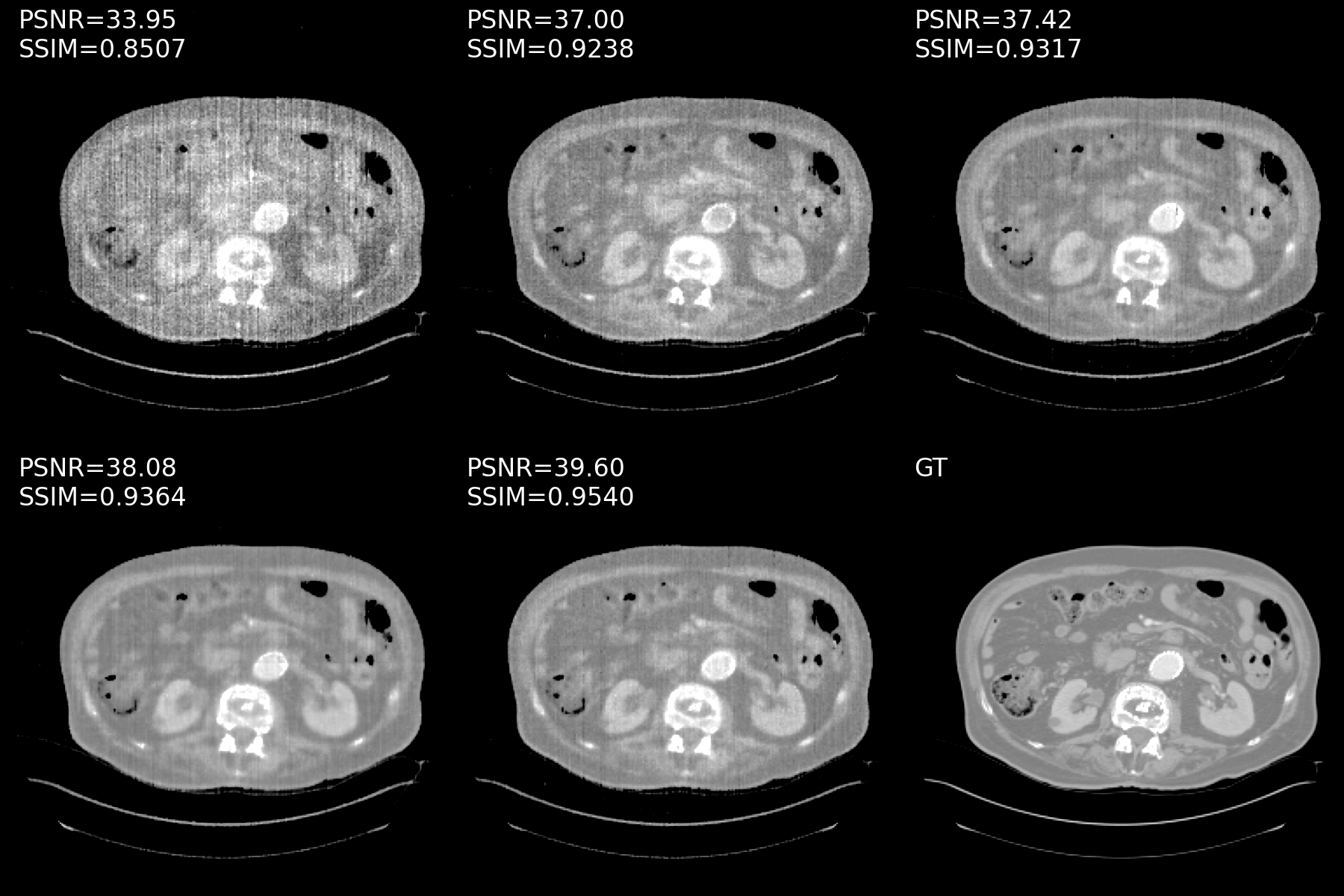} 
  \end{tabular}
  \caption{The SAA phantom slices reconstructed by DIP-ASPINS at different noise levels. From left to right, the noise level at the top row is $I_0 = 10^4, 5\times 10^4, 5\times 10^5.$ In the second row, the noise levels from left to right are $I_0 = 1\times 10^6, 5\times 10^6$, and the ground truth. The display window is $[0.12, 0.35]$.
   }
\label{fig:rec-slices-SAA-diff-noise-level-views-120}
\end{figure*}

The transverse plane image slices reconstructed by DIP-ASPINS at different noise levels are established in \Cref{fig:rec-slices-SAA-diff-noise-level-views-120}. The images in the first row of \Cref{fig:rec-slices-SAA-diff-noise-level-views-120} correspond to the noise level $I_0=10^4, 5\times 10^4, 5\times 10^5$. The second row of \Cref{fig:rec-slices-SAA-diff-noise-level-views-120} shows the reconstructed images of the noisy projection $\bm{Y}$ with the noise level $I_0=10^6,5\times 10^6$ and the ground truth. We can observe that the reconstructed images show improved quality when the noise level is low. Quantitative measurements, such as the PSNR and SSIM values, are marked on the top left of each image. Both the PSNR and SSIM values are increasing along with the decrease in noise level. When the noise level is $I_0=10^4$, severe streak artifacts exist in the reconstructed image slice. More efforts should be made in the future to improve the MSCT reconstruction task at a higher level of noise and sparse view scanning.

\subsubsection{Different SDE Discretization Step $N$}
\begin{figure}[htbp]
  \centering
  \includegraphics[scale=0.435]{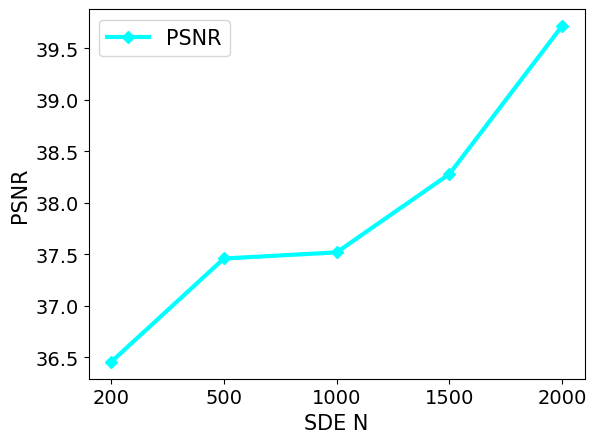}
   \hspace{6mm}
  \includegraphics[scale=0.435]{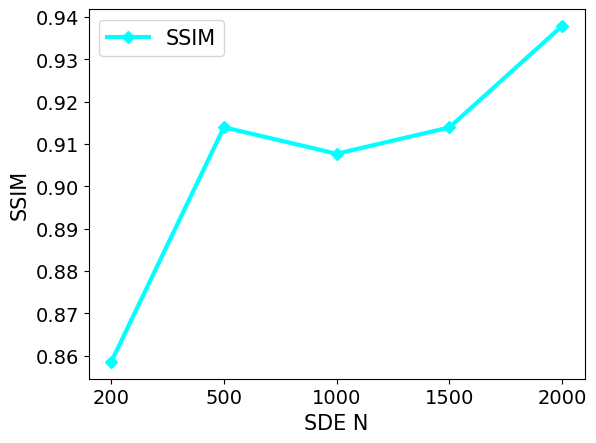}
  \caption{The PSNR (left subfigure) and SSIM (right subfigure) curves with respect to different SDE discretization steps $N$. The test phantom is a 10 slices of Pancreas. The number of sparse views is 72. The test phantom is the Abdomen from the AAPM dataset.}
  \label{fig:diff-sdeN-psnr-ssim}
\end{figure}

The sampling efficiency of the reverse time diffusion process is controlled by the SDE discretization step $N$. We choose a test phantom with 20 slices of the Abdomen and the noiseless projection is simulated. The number of sparse views is set to $\#\mbox{views}=72$. The conditional generation update interval $K$ is chosen based on the value of $N$. When $N$ is larger than 1000, $K$ is set to $50$ within 1000 steps and $K=25$ else. When $N$ is less than 1000, $K$ is set to $25$. The optimization steps in each conditional generation update for INR and SSL are set to $N_{\mbox{INR}}=10$ and $N_{\mbox{SSL}}=50$.
PSNR and SSIM are chosen as metrics to measure the quality of the reconstructed phantom from DIP-ASPINS.
\Cref{fig:diff-sdeN-psnr-ssim} shows the PSNR and SSIM variations with respect to different SDE discretization steps $N=200, 500, 1000, 1500, 2000$. The curve shows that both PSNR and SSIM will increase with the larger SDE discretization step $N$. This leads to the limitation of the proposed DIP-ASPINS model: there should be a balance between the reconstruction quality and the computation time.

\begin{figure*}[htbp]
  \centering
  \begin{tabular}{@{}p{0.98\linewidth}}
    \subfigimg[width=0.98\linewidth]{}{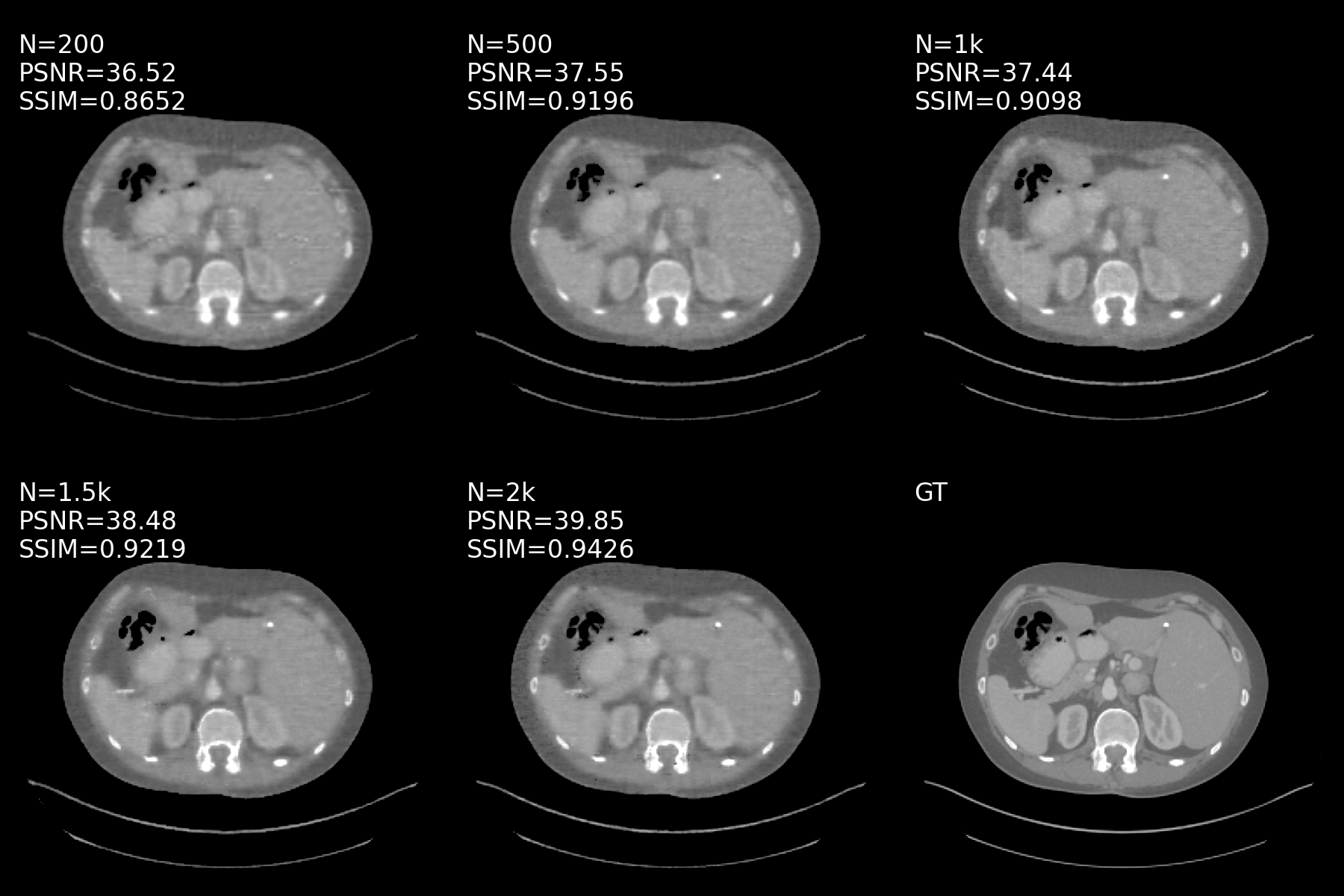} 
  \end{tabular}
  \caption{Reconstructed image slices of the Abdomen phantom by DIP-ASPINS with different SDE discretization step $N$. The values of $N$ at the top row from left to right are $200, 500, and 1000.$ In the second row, the values of $N$ from left to right are set to 1500 and 2000. The last subfigure is the ground truth slice. The display window is $[0.2, 0.5]$.
   }
\label{fig:rec-slices-BE006-diff-sdeN}
\end{figure*}

The image slices reconstructed by DIP-ASPINS at different SDE discretization steps $N$ are shown in \Cref{fig:rec-slices-BE006-diff-sdeN}. In this sparse view imaging setting, the reconstructed image slice has improved quality when the value of $N$ is increased. Streak artifacts appear in the reconstruction slice when the step $N$ is smaller than 1000. Therefore, to obtain a high-quality reconstructed phantom, the SDE discretization step $N$ should be set to a large value, i.e., $N=2000$. The running time of the model will increase along with the reverse time diffusion step $N$. Therefore, one should balance between the image quality and the running time. Quantitative measures (PSNR and SSIM) are marked in the upper left corner of each image slice. It can be seen that both the PSNR and SSIM values are increasing along with the value of $N$ except $N=500$. The little quality measurement gap between the images at $N=500$ and $1000$ does not show distinguishable structural similarity.

\section{Conclusions}\label{sec:conclusions}
In this work, we proposed a diffusion image prior-based model for sparse view multi-source static CT (MSCT) reconstruction. The pre-trained unconditional score function is adopted in the reverse time diffusion process to design a new diffusion posterior sampling strategy incorporating the measurement constraint for a conditional generation. The noisy temporary sample is pushed to noiseless form and then projected onto the affine set to keep the projector consistent with the measured data under different imaging settings. We adopt an implicit neural representation to parameterize the reconstructed phantom to satisfy the practice requirement of high-resolution reconstruction slices. Then, a self-supervised learning model is used to optimize the implicit neural representation model parameters and further enhance the reconstructed image from the conditional diffusion generation process. Numerical experiments verified that the proposed DIP-ASPINS model works well on the multiple static X-ray sources MSCT imaging system at different noise levels, numbers of sparse views, and different SDE generation steps. For future work, we will study the more efficient diffusion posterior sampling scheme to accelerate the conditional generation process. 

\bibliographystyle{siamplain}

\end{document}